\documentclass[12pt]{JHEP3}
\usepackage{epsfig}

\makeatletter
\@addtoreset{equation}{section}

\preprint{
KIAS-P03033\\
hep-th/0305229\\
}

\title{\Large\bf String Fluid, Tachyon Matter,
and Domain Walls}

\author{O-Kab Kwon$^{a,b}$ and Piljin  Yi$^b$\\ \\
\llap{$^a$} BK21 Physics Research Division and Institute of Basic Science\\
Sungkyunkwan University, Suwon, Korea\\ \\
\llap{$^b$} School of Physics, Korea Institute for Advanced Study\\
207-43, Cheongryangri-Dong, Dongdaemun-Gu, Seoul 130-012, Korea\\
\\
E-mail: \email{okwon@newton.skku.ac.kr}, \email{piljin@kias.re.kr}}

\abstract{\vskip 5mm
We study classical dynamics of an open string tachyon $T$ of unstable D$p$-brane
coupled to the gauge field $A_\mu$. In the vacuum with vanishing potential,
$V=0$, two fluid-like degrees of freedom, string fluid and tachyon matter,
survive the tachyon condensation. We offer general analysis of the
associated Hamiltonian dynamics in arbitrary background. The canonical field
equations are organized into two sets, fluid equations of motion augmented by
an integrability condition. We show that a large class of motionless
and degenerate family of classical solutions exist and represent arbitrary
transverse distribution of tachyon matter and flux lines. We further test their
stability by perturbing the fluid equation up to the second order. 
\vskip 5mm
Second half of this note considers possibility of $V \neq 0$ in the dynamics. 
We incorporate $V$ in the Hamiltonian equation of motion and consider
interaction between domain walls and string fluid. During initial phase of 
tachyon condensation, topological defect at $T=0$  is shown to attract 
nearby and parallel flux lines. The final state is fundamental 
strings absorbed and spread in some singular D$(p-1)$ brane soliton. When string
fluid is transverse to the domain wall, the latter is known to turn into a smooth
solution. We point out that a minimal solution of this sort exists
and saturates a BPS energy bound of fundamental string ending on a D$(p-1)$ brane.}

\begin{document}

\pagebreak
\renewcommand{\thepage}{\arabic{page}}
\tableofcontents
\pagebreak

\section{String Fluid and Tachyon}

Decay of unstable D-branes \cite{Sen:1998sm} have served as useful
laboratories of understanding off-shell structure of string theory.
After several years of intense study, we seem to have fairly good idea how
lower dimensional, stable D-branes are formed out of decaying unstable
D-branes. They arises as topological defects, familiar in ordinary
field theories, although solutions tends to be a bit singular than
usual \cite{Kraus:2000nj,Takayanagi:2000rz, Sen:2003tm}.
Among very well understood are how the relevant space-time
Ramond-Ramond charges are generated via winding numbers and
topology of gauge bundles \cite{Witten:1998cd}.

One of reasons that we should anticipate lower dimensional
D-branes to emerge at the end of day is the charge conservation.
Stable D-branes carry Ramond-Ramond gauge charge
\cite{Polchinski:1995mt},
which can be
absorbed in unstable D-brane and transmutes into world-volume gauge
field configurations \cite{Strominger:1995ac}. Charge conservation
in the space-time begs for a
mechanism for recovering these conserved quantum numbers, and the
only objects that can carry such charge after the unstable brane
annihilate itself, would be the stable D-branes themselves. Thus,
we must somehow be able to reproduce the D-branes within the
dynamics of unstable D-branes.

One would expect the same logic should apply to fundamental strings
\cite{Yi:1999hd, Bergman:2000xf},
say, to infinitely long semiclassical ones at least. Yet, emergence
of fundamental strings has proved a much more challenging problem.
Some tantalizing hints have been accumulating via study of low energy
effective action of D-brane decay, nevertheless, and in this note,
we will give a comprehensive review of the classical low energy dynamics
with a commonly employed effective action, conceived by
many authors, and study various aspects
with a view toward formation of fundamental string.
The form of the Lagrangian we will
study is of the form \cite{Sen:1999xm,Sen:1999md,
Garousi:2000tr, Bergshoeff:2000dq,
Kluson:2000iy,Arutyunov:2000pe,Kutasov:2003er},
\begin{equation}
-V(T)\sqrt{-Det(\eta+F+\nabla X^I \nabla X^I +\nabla T\nabla T)},
\end{equation}
where the tachyon $T$ appears inside the determinant on equal footing
as the transverse scalars $X^I$. $V(T)$ is everywhere nonnegative and has
a runaway behavior \cite{Gerasimov:2000zp,Kutasov:2000qp,Kutasov:2000aq}.
That is $V(T)$ vanishes exponentially at
$T=\pm \infty$. For first half of this note, we will consider $V=0$
strictly. Latter part of the note will incorporate $V\neq 0$ and compute
its effect near domain wall formation.

The earliest evidence that this low energy field theory is quite
unconventional came in Ref.~\cite{Bergman:2000xf},
which considered pure gauge
dynamics in decay of unstable D2-brane. They found that the electric
flux lines become free in that no transverse pressure is present and
also that the tension density of the flux lines obeys a BPS-like property
which would be normally seen in fundamental string.
This degenerate behavior of flux lines were later found to persist
in higher dimensional case as well
\cite{Gibbons:2000hf}, once $V(T)$ is taken to vanish.

This pressureless collection of flux lines were dubbed ``string fluid,''
in obvious reference to the fact that they carry fundamental string
charges.\footnote{This fluid-like behavior had been also noticed  as
the strong coupling limit of Born-Infeld system \cite{Lindstrom:1997uj,
Gustafsson:1998ej,Lindstrom:1999tk}.}
One outstanding question is whether and how properly quantized
fundamental string emerges from this string fluid. Tantalizingly similarities
already exist between string fluid  and classical fundamental string
raising hope that via some confinement mechanism we may recover fundamental
string from decay of unstable D-branes.
Classical properties of string fluid are studied in detail in
Ref.~\cite{Gibbons:2000hf}.

It turns out such fluid-like behavior is not limited to the gauge sector
but extends to the tachyon. When the final state is static and homogeneous,
an on-shell condition reads \cite{Gibbons:2000hf,Sen:2002nu,Gibbons:2002tv}
\begin{equation}
1=E_iE_i + \dot T^2,
\end{equation}
where $E_i=F_{0i}$ are components of the field strength.
Energy density of any motionless state is composed of two components
\begin{equation}
H_{motionless} =\sqrt{\pi^i\pi^i +\pi_T^2},
\end{equation}
where $\pi^i$ is the conserved electric flux, such that $\nabla_i\pi^i=0$,
while $\pi_T$ is the canonical momenta of $T$. The kinematics are
such that we have a relationship,
\begin{equation}
{\pi_i}/{\pi_T}={E}_i/{\dot T}.
\end{equation}
Pure string fluid emerges when $\dot T=0=\pi_T$, while the other
limit $\vec E=0$ involves energy density composed solely of $\pi_T$.
In the latter limit, all energy is carried by dust-like matter,
known as tachyon matter \cite{Sen:2002in,Sen:2002an}.

One of less understood aspects of the combined system of string fluid and
the tachyon matter is how the two components interact with each other
\cite{Gibbons:2002tv, Ishida:2002fr, Mehen:2002xr, Kim:2003he}
This system was addressed comprehensively and without any
approximation, in Ref.~\cite{Gibbons:2002tv}.
Among explained are the origin of the pressureless nature of these
fluid, and also exact Hamiltonian, canonical field equations, and
energy-momentum tensor are written down. The combined fluid shows
character of 1+1 dimensional system with variable ``speed of light",
which depends on the composition of the two fluid. Although exact
``static" solutions were found, generic aspect of the dynamics has
been poorly understood. In this note, we will address this issue and
try to understand how one fluid component react to the presence of the
other.

We provide a comprehensive analysis of the classical
dynamics in the Hamiltonian formulation, taking into account possible
coupling to background metric. One purpose is to clarify the relation
between the fluid equation of motions and the Hamiltonian equation
of motion. Previously, the former has been derived from combination of
the latter and the energy-momentum conservation. In this note, we will
describe in what sense solution to the fluid equation gives solutions to
the field equations. Sections 2,3, and 4 are devoted to this.

In particular, this allows us a rather sweeping
characterization of all static (meaning that physical momenta vanishes
identically) classical solutions and shows their rather huge degeneracy.
After isolating such static solutions, we test their stability by
perturbing the fluid system in section 5.
Perturbation up to 2nd order is performed,
which looks pretty involved, and we argue that all physical effect
from such perturbation may be understood as a consequence of continuity
of the fluid and lacks any destabilizing interaction.

In fact, the aggregation of energy density and flux lines turns
out to be tied to formation of domain walls, instead. In section 6, 
we extend the Hamiltonian dynamics formulation to include
possibility $V\neq 0$. We show that the domain-wall tends
to attract nearby and parallel flux lines via a short range attraction.
The effective range of this attraction is given by region with finite $V$,
and collapses as stable D$(p-1)$ branes form. However, once the domain
wall formation is complete, the range of attractive interaction become
arbitrarily small, and such aggregation of (parallel) string fluid is
no longer favored.\footnote{We must caution readers that this effect 
is unrelated to usual attraction between D-branes and 
parallel fundamental string. The latter arises from exchange of closed 
strings between the two objects, and thus is normally associated with 
open string one-loop. The net effect is the same, nevertheless, so such a
tendency to form bound state of a D$(p-1)$ and fundamental strings is not 
too surprising.}

In section 7, we consider another configuration involving string
fluid and domain wall, where the former lies  orthogonal to the
latter. Smooth solutions of this kind were recently written down, and 
here we reproduce them from our Hamiltonian description
above. In particular, we show that a minimal  solution
exists and saturate BPS bound which is normally associated with
1/4 BPS configuration of fundamental string(s) ending on or 
passing through a D-brane. We close with a summary.

\section{Fluid Equation, Integrability Condition, and Classical Solutions}

Fluid equation for $V=0$ was first written down in Ref.~\cite{Gibbons:2000hf}.
The main purpose there was to understand gauge dynamics, but the analysis
did deal with possibility of turning on transverse scalars and also dynamical
$T$ provided that the kinetic term of the latter shows up on equal
footing as transverse scalars $X^I$. This is precisely the Lagrangian
in question. Furthermore, any scalar whose kinetic term appears this
way can be treated as if it is a component of gauge field along some
hidden direction, and manipulations for gauge field carries over almost
verbatim. Because of this, the result carries over immediately
to the system of string fluid coupled to tachyon matter. In this section,
we will rewrite the result with tachyon field explicitly expressed,
and consider its implications.
In section 3 and 4, we will elevate this to the general background with
curved metric.

The main idea is to deal with the dynamics from the Hamiltonian viewpoint.
Let us introduce extended gauge field $A_M=(A_\mu,A_T)$ so that
$A_T\equiv T$. Then the Lagrangian of tachyon coupled to a gauge field
may be written succinctly as \cite{Gibbons:1997xz}
\begin{equation}
-V(T)\sqrt{-{\rm det}(\eta_{MN}+F_{MN})},
\end{equation}
where $F_{MN}$ is the field strength associate with $A_M$ with the
understanding that $\partial_T\equiv 0$. Since we will be using
Hamiltonian formulation we will be separating out time direction
from the ``spatial" ones. For the latter we will use subscript
$m=1,\cdots,p,T$ and $i=1,\cdots, p$.
Using Legendre transformation, we obtain the following
Hamiltonian of this system
\begin{equation}\label{genH}
{\cal H} = H-A_0 \partial_i \pi^i,
\end{equation}
where
\begin{equation}
H= \sqrt{\pi^m \pi^m
                  + P_m P_m + V^2 \det (\delta_{mn}+F_{mn})}.
\end{equation}
Here $\pi^m$ is the canonical conjugate
momenta, defined as $ \partial {\cal L}/\partial \dot A_m$,
while $P_m \equiv \pi^n F_{mn}$. $P_i$ with $i=1,2,\cdots, p$
correspond   to the conserved momenta associated with the
translational symmetry.

In section 2 through
section 5, we will consider the limit, $V=0$, which represent
the final stage of tachyon condensation where tachyon $T$ is
rolling away to $\infty$ everywhere. The Hamiltonian (in temporal gauge)
is exceedingly simple, and has
\begin{equation}
{H}= \sqrt{\pi^m\pi^m+P_mP_m}.
\end{equation}
Note that despite the vanishing
Lagrangian as $V\rightarrow 0$, the Hamiltonian remains finite.

\subsection{Fluid Equations}

Half of the canonical equation of motion
\begin{equation}
\dot \pi^m=-\frac{\delta}{\delta A_m} \int dx^p \;{\cal H},
\end{equation}
combined with conservation of
energy momentum gives immediately the following fluid equations
\cite{Gibbons:2000hf}
\begin{eqnarray}
\partial_0 n^m +v^i\partial_i n^m=n^i\partial_i v^m, &&\nonumber \\
\partial_0 v^m +v^i\partial_i v^m=n^i\partial_i n^m, &&\label{eof}
\end{eqnarray}
where the vector fields $n$ and $v$ are defined as
\begin{eqnarray}
\pi={ H}n, &&\nonumber \\
{ P}={H}v,
\end{eqnarray}
which satisfies the constraints
\begin{equation}
n^mn^m+v^mv^m=1,\qquad n^mv^m=0.
\end{equation}
The evolution of the energy density ${H}$ is then determined via,
\begin{equation}
\partial_0{H}+\partial_i\left({ H}v^i\right)=0,
\end{equation}
and finally $\pi^i$ must satisfy Gauss's constraint,
\begin{equation}
\partial_i \pi^i=0.
\end{equation}
The Noether momenta $P_i$ has the following simple expression
\begin{equation}
P_i=-F_{ij}\pi^j-\partial_iT \pi_T=\pi_m F_{mi}.
\end{equation}
The last component of the vector $P$, $P_T$, is not really a
conserved momenta since $T$-th direction is a mere mathematical
device of convenience. It is nevertheless computed by pretending
$T$-th direction exists as a translationally invariant
spatial direction, and thus given by the combination
\begin{equation}
P_T\equiv \partial_i T\pi^i=\pi_mF_{mT}.
\end{equation}
Most of quantities here have simple physical interpretation.
$\pi^i$ is nothing but conserved electric flux, while $\pi_T$ is the tachyon
matter density. $ H$ and $P_i$ are conserved energy and
momentum density, respectively. The last quantity $P_T$ measure inhomogeneity
of tachyon $T$ along the flux direction. The fact that
this appears in the Hamiltonian separately implies a rather
anisotropic behavior the system.

\subsection{Integrability Conditions}

Although fluid equations are self-contained, a solution may
not give automatically a solution to the original field
equations. This is because that the fluid variables above are
naturally formed from canonical variables and does not produce
elementary fields $A_m$ directly. A further set of first-order
equations must be solved. These arise from
\begin{equation}
\dot A_n=\frac{\delta }{\delta \pi^n}\int dx^p \;{\cal H},
\end{equation}
and are exactly half of the Hamiltonian equations of motion.

These equations are intimately related to the fact that the determinant
part of the Lagrangian vanishes. The Hamiltonian with $V$ kept  is
such that
\begin{equation}
H=\sqrt{\pi_m C_{mn} \pi_n +O(V^2)},
\end{equation}
for some matrix $C$ independent of $\pi_m$. Then the Lagrangian
is
\begin{equation}
L=\dot A_m\pi_m- {\cal H}
=\pi_m\left(\frac{\delta}{\delta \pi^m} \int {\cal H}\right)
-{\cal H} =O(V^2).
\end{equation}
On the other hand the Lagrangian is of the form
\begin{equation}
-V\sqrt{-{\rm det}(\cdots)},
\end{equation}
so the on-shell value of Lagrangian must be such that
\begin{equation}
\sqrt{-{\rm det}(\cdots)}\sim V\rightarrow 0,
\end{equation}
which is precisely the condition that generalizes,
\begin{equation}
1-\dot T^2-E_iE_i \rightarrow 0 ,
\end{equation}
of the homogeneously rolling tachyon.

In any case, these equations
can be expressed as an algebraic equation for the field
strength  associated with elementary fields
$A_q$ in terms of fluid variables, $H$, $n$, $v$;
\begin{equation}
E_m=n_m+F_{mn}v_n\label{alg},
\end{equation}
where $E_m=F_{0p}$. This, combined with  the Bianchi identity
\begin{equation}
\dot F_{mn}=\partial_m E_n -\partial_n E_m ,
\end{equation}
gives an evolution equation for  $F_{mn}$.

On the surface, thus,  it may seem that we have split the system
into two steps; solve the first-order fluid equation
and then solve another first order equations for elementary fields
in the background of $H$, $ n$, and $ v$. However, there is
one subtlety here. The two fluid variables $ v$ and
$n$ are algebraically related as
\begin{equation}
v_m=n_n F_{nm}, \label{int}
\end{equation}
and there is a logical possibility that one may not find such
$F_{mn}$ as a solution. Only if there is a solution $F_{mn}$
consistent with (\ref{int}), the solution to fluid equation
would be acceptable.

In the simplest case of pure tachyon, this integrability condition
appears in a particularly simple manner. The only elementary field
is $A_T=T$ in that case, and we have $n_i=0$. Then
(\ref{int}) states that
\begin{equation}
v_i=-\partial_i T n_T=\mp\frac{\partial_i T}{\sqrt{1+(\partial_i T)^2}},
\end{equation}
while (\ref{alg}) gives,
\begin{equation}
\dot T = \left(1+(\partial_i T)^2\right) n_T =\pm\sqrt{1+(\partial_i T)^2}.
\end{equation}
In this case the
integrability condition may be imposed as a familiar local condition
on the fluid system as
\begin{equation}
\partial_i \left(\frac{v_j}{n_T}\right)- \partial_j
\left(\frac{v_i}{n_T}\right) =0.
\end{equation}
In section 3, we will further consider the pure tachyon case and reformulate
these conditions in a manifestly relativistic manner.

\subsection{``Static" Solutions}

Let us characterize all static solutions. By ``static" we mean absence
of physical momentum, $P_i=0$, and this immediately constrains
us to look for configurations with $v_i=0$. One of the algebraic
constraint then implies that $n^Tv_T=0$. This can be achieved
by $v_T=0$ in addition to $v_i=0$, and if we choose this,\footnote{The other
choice of $n_T=0$ leads to a different class of solution that involves
domain walls \cite{Kim:2003in,0305092,Brax:2003rs}.
Since $0\neq v_T=n_i\partial_i T$ implies nonvanishing
gradient of $T$ along the flux direction, such a configuration will gave
$T=0$ somewhere along the flux line direction. The final state would
corresponds to a domain wall threaded by a transverse string fluid.
Because of this, one
can no longer restrict to $V\rightarrow 0$ limit, and one must solve
for the full equation of motion. See section 7
 for this class of solutions.}
fluid equation collapse to
\begin{eqnarray}
\partial_0{H}&=&0, \\
\partial_0 n^m&=&0,
\end{eqnarray}
and
\begin{eqnarray}
n^i\partial_i n^m=0,&& \label{straight} \\
\partial_i\left({H}n^i\right)=0.&&
\end{eqnarray}
Because of (\ref{straight}),
flux lines should  be straight, and we may as well
associate its direction with $x^1$ while denoting the remaining
$p-1$ directions by $x^{2,3,\cdots,p}$.
The most general ``static" solutions (with $v_T=0$) to the
fluid equations are time-independent and $x^1$-independent
distribution of flux $\pi^1$ and tachyon matter $\pi^T$.
By construction the same property holds for ${H}$. Then, all
``static" classical solutions are characterized by the following
two arbitrary and independent (nonnegative)
functions\footnote{An exceptional case is when $\pi^i$ also
happens to vanish; $\pi^T$ could  then be an arbitrary function of
$x^{1,2,3,\cdots,p}$. }
\begin{eqnarray}
\pi^1(x^2,x^3,\cdots, x^p),&&\nonumber \\
\pi^T(x^2,x^3,\cdots, x^p) ,&&\nonumber
\end{eqnarray}
together with the choice of $x^1$ direction or equivalently
the choice of the electric flux direction. The energy density
is
\begin{equation}
H(x^2,\cdots, x^p)=\sqrt{\left(\pi^1(x^2,x^3,\cdots, x^p)\right)^2+
\left(\pi^T(x^2,x^3,\cdots, x^p)\right)^2}.
\end{equation}
In section 5, we will analyze
classical stability of this large family of static solutions.

We still need to test whether this ``static" solution satisfies all the
integrability condition. Namely, we need to find a gauge field, $A_m$,
solving (\ref{alg}),
such that (\ref{int}) holds. Since $\vec v=0$ by construction, we have
\begin{equation}
E_m=n_m=\pi_m/H,
\end{equation}
which has a solution in the temporal gauge $A_0=0$
\begin{equation}
A_m=t n_m (x^2,\cdots, x^p).
\end{equation}
Then the magnetic field is
\begin{eqnarray}
-F_{1j}=F_{j\,1}&=& t \partial_j n_1 ,\nonumber \\
-F_{Tj}=F_{jT}&=& t \partial_j n_T  ,\nonumber
\end{eqnarray}
for $j=2,3,\cdots,p$. All other components vanish. Testing whether
this gives back a vanishing $v$ via $v_m=n_n F_{nm}$, we
find
\begin{eqnarray}
&&v_1=0, \nonumber \\
&&v_j=-F_{j\,1}n_1-F_{j\, T} n_T=
-t\partial_j\left(\frac{n_1^2+n_T^2}{2}\right)=0, \nonumber \\
&& v_T=0.
\end{eqnarray}
Thus our ``static" solution is integrable and thus
acceptable as a solution to the original field equations.

\section{Tachyon Matter in General Background}

The tachyon effective action on an unstable D$p$ brane in
general gravity background is given by
\begin{equation}\label{pta}
S= -\int d^{p+1} x V(T) \sqrt{-g} \sqrt{1 + g^{\mu\nu}
    \partial_\mu T \partial_\nu T},
\end{equation}
where $g \equiv \det g_{\mu\nu}$ and $V(T)$ is the tachyon potential,
and we defined the general metric as
\begin{eqnarray}\label{metr}
ds^2 &=& g_{\mu\nu} dx^\mu dx^\nu
      = - N^2 dt^2 +h_{ij}(dx^i + L^i dt)(dx^j + L^j dt)
\nonumber \\
     &=& g^{\mu\nu} \partial_\mu\partial_\nu
      = -\frac{1}{N^2} (\partial_t - L^i\partial_i)^2
        + h^{ij} \partial_i\partial_j,
\qquad (1 \leq \ i,j \ \leq p).
\end{eqnarray}
This general metric is determined
by the lapse function $N$, shift vector $L^i$,
and spatial metric $h_{ij}$ of the $p$-dimensional
hyper-surface at a given time slice.
Using this general metric (\ref{metr}), we can rewrite the action
(\ref{pta}) as
\begin{eqnarray}\label{pta2}
S &=& \int dt \int d^p x \,{\cal L},
\nonumber \\
{\cal L} &=& - V(T) N\sqrt{h}\sqrt{\tilde X},
\end{eqnarray}
with $h \equiv \det h_{ij}$, $g = -N^2 h$  and
\begin{eqnarray}\label{tX}
\tilde X &\equiv& 1 + g^{\mu\nu}\partial_\mu T \partial_\nu T
\nonumber \\
          &=& 1 - \frac1{N^2}(\dot T - L^i \partial_i T)^2
                + h^{ij} \partial_i T\partial_j T,
\end{eqnarray}
where the doted notation denotes time derivative.
Then the Hamiltonian is obtained by
\begin{eqnarray}\label{pham}
{\cal H} &=& \pi \dot T - {\cal L}
\nonumber \\
         &=& \pi\dot T + V(T) N\sqrt{h}\sqrt{\tilde X}
\nonumber \\
         &=& N\sqrt{\pi^2 + (\pi \partial_i T) h^{ij}
               (\pi \partial_i T) + V^2 h (1+ h^{ij}
                \partial_i T \partial_j T)}
                + \pi L^i \partial_i T,
\end{eqnarray}
in which we have defined the conjugate momentum as
\begin{equation}
\pi \equiv \frac{\partial {\cal L}}{\partial \dot T}
         = \frac{\sqrt{h}V}{N\sqrt{\tilde X}}
           (\dot T - L^i \partial_i T).
\end{equation}
After tachyon condensation, i.e., in $V\rightarrow 0$ limit,
the Hamiltonian equations are given by
\begin{eqnarray}
\dot T &=& \frac{\partial {\cal H}}{\partial \pi}
        = N\sqrt{1 + h^{ij}\partial_i T\partial_j T}
          + L^i \partial_i T,
\label{pheq1} \\
\dot \pi &=& -\frac{\partial {\cal H}}{\partial T}
          = \partial_i \left( \frac{N\pi h^{ij} \partial_j T}{
            \sqrt{1 + h^{kl}\partial_k T\partial_l T}}\right)
            + \partial_i(\pi L^i).
\label{pheq2}
\end{eqnarray}
Let us consider the following Lorentz invariant matter density
\begin{eqnarray}\label{inv}
\mu \equiv \frac{p^\alpha\partial_\alpha T}{\sqrt{-g}}
      &=& \frac{V}{\sqrt{\tilde X}}
           \left( \frac1{N^2}(\dot T - L^i\partial_i T)^2
           - h^{ij}\partial_i T \partial_j T\right)
\nonumber \\
     &=&\frac{V}{\sqrt{\tilde X}},
\end{eqnarray}
where we define
\begin{equation}
p^\alpha \equiv
\frac{\partial {\cal L}}{\partial (\partial_\alpha T)},
\end{equation}
so that $p^0=\pi_T$.
In the last step of the above equation, we used the Eq.~(\ref{pheq1}).

Using the equations (\ref{pheq1}), (\ref{pheq2}),
we obtain
\begin{eqnarray}\label{gdeq1}
\nabla_\mu(\mu \nabla^\mu T)
       &=& \frac1{\sqrt{-g}} \partial_\mu(\mu \sqrt{-g}
           g^{\mu\nu}\partial_\nu T)
\nonumber \\
      &=& \frac1{\sqrt{-g}} \partial_0\left[
          \mu\sqrt{-g}\left(-\frac1{N^2}~\dot T
          + \frac{L^i}{N^2}
            \partial_i T\right)\right]
\nonumber \\
        &&+ \frac1{\sqrt{-g}} \partial_i
           \left[\mu \sqrt{-g}\left(\frac{L^i}{N^2} \dot T
           +h^{ij}\partial_j T
           - \frac{L^iL^j}{N^2}\partial_j T\right)\right]
\nonumber \\
      &=& \frac1{\sqrt{-g}}\left[ -\partial_0 \pi
          +\partial_i(\pi L^i) + \partial_i\left(
          \frac{N\sqrt{h} V}{\sqrt{\tilde X}} h^{ij}
          \partial_j T\right) \right] = 0.
\end{eqnarray}
We used the Eq.~(\ref{pheq2}) in the last step of the Eq.~(\ref{gdeq1}).
Then the energy momentum conservation implies that
\begin{eqnarray}
0=\nabla^\mu T_{\mu\nu}&=&\nabla^\mu \left(\frac{V}{\sqrt{\tilde X}}
               \partial_\mu T \partial_\nu T\right) \nonumber \\
               &=&\mu
               \partial_\mu T \nabla^\mu \partial_\nu T. \label{cons}
\end{eqnarray}
All of these have a rather obvious interpretation once we identify
$\mu$ as the invariant matter density and $-\partial_\mu T$ as the
velocity field $U_\mu$ \cite{Sen:2002qa}.
Equations (\ref{gdeq1}) and (\ref{cons}) then
implies,
\begin{eqnarray}
\nabla_\alpha \left(\mu U^\alpha\right)&=& 0,\\
U^\alpha\nabla_\alpha U^\mu&=&0, \label{free}
\end{eqnarray}
which are nothing but the continuity equation and the geodesic equation
for dust. These are the two fluid equation for case of pure tachyon.

If we were treating the density $\mu$ and velocity field $U_\mu$ as
elementary quantities, the fluid equations (\ref{free}) must be
augmented by an integrability condition
\begin{equation}
\nabla_\mu U_\nu-\nabla_\nu U_\mu=0,
\end{equation}
to make contact with the original field theory. This of course gives
$U_\mu =-\partial_\mu T$ for some function $T$, and we would interpret this
$T$ as the original tachyon of the system.
Also recoverable from the field equations is the fact that
\begin{equation} \label{conlim}
U_\mu  U^\mu  = -1,
\end{equation}
which is a kinematical constraint, saying that the velocity field
$U_\mu=-\partial_\mu T$ arises from affine parameterization of the
geodesics. Obviously it does not affect motion of the tachyon matter,
and simply gives how $T$ should be solved for, given any particular
set of trajectories of tachyon matter.

Thus the tachyon matter in the tachyon condensation limit corresponds
with an ideal (rotationless) fluid, moving freely along geodesics,
with the  trajectory being affine-parameterized.  In particular, this
implies that the tachyon matter clusters under the gravitational
interactions just as ordinary matter does.

\section{String Fluid Coupled to Tachyon Matter}

Here we repeat the above analysis with gauge field included.
The fluid equation for string fluid were first written in
Ref.~\cite{Gibbons:2000hf}, and here we show their completeness again in that
all nontrivial dynamical information can be recovered just
from solving the fluid equations.
The effective action for tachyon coupled to Abelian gauge field
on D$p$-brane can be written by
\begin{eqnarray} \label{ga}
S &=& \int d^{p+1} x {\cal L}, \\
{\cal L} &=& -V(T) \sqrt{-X}
\nonumber
\end{eqnarray}
with
\begin{eqnarray}
X &\equiv& \det (g_{\mu\nu}
      + \partial_\mu T \partial_\nu T + F_{\mu\nu}),
        \qquad (\mu, \nu = 0,1, \cdots, p),
\label{detX} \\
F &=& \partial_\mu A_\nu - \partial_\nu A_\mu,
\nonumber
\end{eqnarray}
where $g_{\mu\nu}$ is defined in Eq.~(\ref{metr}).

In calculating the determinant in Eq.~(\ref{detX}),
it is convenient to consider the tachyon field $T$
as $(p+1)$-component of the gauge field, as we already
used in section 2. The metric is also extended to
include a flat fictitious direction $x^T$.
Then the Eq.~(\ref{detX}) can be rewritten by
\begin{eqnarray}\label{genX}
X &=& \det (g_{MN} + F_{MN}),\qquad (M,N = 0,1,\cdots, p, T),
\end{eqnarray}
where $F_{\mu T} = \partial_\mu T$ since $\partial_T =0$.
Using Legendre transformation, we obtain the following
Hamiltonian of this system
\begin{equation}
{\cal H} = N \sqrt{\pi^m h_{mn} \pi^n
                  + P_m h^{mn}P_n + V^2 \det (X_{mn})}
           -P_m L^m - A_0\partial_i \pi^i,
\end{equation}
where $\pi^m \equiv \delta {\cal L}/\delta
\dot A_m$, $P_m \equiv \pi^n F_{nm}$. Canonical field equations
for this system can be found in Ref.~\cite{Gibbons:2002tv}, and in
the following we will reexpress these in terms of fluid-like
variables and also generalize it to include nontrivial
background metric.

\subsection{Generalities: Case of an Unstable D2-Brane}

Our purpose in this paper is to understand the interaction
between tachyon matter and electric flux line.
To accomplish this purpose, we concentrate on the simplest case,
D$2$-brane.
We believe that the tachyon condensation on D$2$-brane contains all
nontrivial characteristics of the interaction of tachyon and
flux in general D$p$-brane decay.
The determinant in Eq.~(\ref{detX}) on D$2$-brane system is
given by
\begin{equation}
X = g\left(1 + \frac12 F^2
    - \frac1{16} ({F^*} F)^2\right)
  = g \tilde X,
\end{equation}
with
\begin{eqnarray}
F^2 =  F_{MN} F^{MN},
\qquad
F^* F = {F^*}_{MN} F^{MN},
\qquad
{F^*}^{MN} = \frac1{2\sqrt{-g}} \epsilon^{MNPQ} F_{PQ},
\end{eqnarray}
where $M,N = 0,1,2,T$ and $\epsilon^{MNPQ}$ is the Levi-Civita symbol and
we choose $\epsilon_{012T} = 1$, $\epsilon^{012T} = -1$.
Energy-momentum tensor $T_{\mu\nu}$ is\footnote{Bear in mind that, since
we artificially introduced $x^T$ direction, there is no conserved momentum
associated with this direction.}
\begin{eqnarray}\label{emt}
T_{\mu\nu} \equiv -\frac{2}{\sqrt{-g}} \frac{\delta S}{\delta g^{\mu\nu}}
          = -\frac{V}{\sqrt{\tilde X}} \frac{C^S_{\mu\nu}}{g},
\end{eqnarray}
where $C^S_{MN}$ is the symmetric part of the cofactor $C_{MN}$ for
$X_{MN} = g_{MN} + F_{MN}$ which is given by
\begin{equation}\label{cf}
C^{MN} = g\left(g^{MN} (1 + \frac12 F^2) + F^{MN} + F^{MP}
         {F_P}^{N} -\frac14 {F^*}^{MN} (F^*F)\right).
\end{equation}
Equation of motion for the gauge field $A_M$ and
conservation of energy-momentum are expressed by
\begin{eqnarray}\label{geom}
&&\nabla_\mu\left(\frac{V}{\sqrt{\tilde X}}
          \frac{C^{\mu N}_A}{g}\right) = 0,
\\
&& \nabla_\mu\left( \frac{V}{\sqrt{\tilde X}}
            \frac{C^{\mu\nu}_S}{g} \right) = 0,
\label{cem}
\end{eqnarray}
where $C_A^{MN}$ is the anti-symmetric part of the cofactor
in Eq.~(\ref{cf}).
Conjugate momentum for gauge field and conserved Noether momentum which is
induced by invariance of the spatial translation are given by
\begin{eqnarray}
\pi^m =\frac{V}{\sqrt{\tilde X}} \frac{C^{0m}_A}{\sqrt{-g}},
\qquad
P_i = \frac{V}{\sqrt{\tilde X}}
      \frac{C^{0m}_A}{\sqrt{-g}} F_{mi}.
\label{gmom}
\end{eqnarray}
In the next sections, we will do that in tachyon
condensation limit in flat and curved space.

\subsection{Fluid Equations in Flat Space}

In flat space metric $\eta_{MN} = \mbox{diag}(-1,1,1,1)$,
the Hamiltonian density (\ref{genH})
in tachyon condensation limit ($V=0$) is given by
\begin{equation}
H = \sqrt{{\vec\pi}^2
              + (\vec B \times \vec \pi)^2}
         = \frac{V}{\sqrt{\tilde X}}
               (1 + {\vec B}^2),
\label{fH}
\end{equation}
where we use $A_0=0$ gauge and define $E_m\equiv F_{0m}$,
$B_m \equiv \epsilon_{mnl} F_{nl}$ and thus
\begin{equation}\label{fX}
\tilde X = 1 + {\vec B}^2 - {\vec E}^2
             - (\vec E \cdot \vec B)^2 = 0 ,
\end{equation}
with an explicit expression
\begin{eqnarray}
\vec E =(E_1, E_2, \dot T), \qquad
\vec B = (\partial_2 T, - \partial_1T, B).
\label{EB}
\end{eqnarray}
With this, we find
\begin{equation}
\pi_m = \frac{V}{\sqrt{\tilde X}}
          \left(E_m + B_m (\vec E \cdot \vec B)\right),
\qquad
P_m = (\vec B \times \vec \pi)_m =
      \frac{V}{\sqrt{\tilde X}} \epsilon_{mnl} B_n E_l,
\label{fmom}
\end{equation}
and under the condition (\ref{fX}),
\begin{eqnarray}
\frac{V}{\sqrt{\tilde X}} C^{mn}_A &=&
\frac{P_m\pi_n - \pi_m P_n}{ H},
\nonumber \\
\frac{V}{\sqrt{\tilde X}}C^{ij}_S &=&
           \frac{P_i P_j - \pi_i \pi_j}{ H}.
\label{fcof}
\end{eqnarray}
Equations of motion gives
\begin{eqnarray}
&&\vec\partial\cdot ( H \vec n) = 0,
\label{nv1} \\
&&\dot{\vec n} + (\vec v\cdot \vec\partial)\vec n =
             (\vec n \cdot \vec\partial) \vec v,
\label{nv3}
\end{eqnarray}
when augmented with the energy conservation,
\begin{eqnarray}
&& \dot  H + \vec\partial\cdot (  H \vec v) = 0.
\label{nv2}
\end{eqnarray}
The momentum conservation may be written as
\begin{eqnarray}
&&\dot{\vec v} + (\vec v\cdot \vec\partial)\vec v =
            (\vec n \cdot \vec\partial)\vec n,
\label{nv4}
\end{eqnarray}
where energy conservation is taken into account. Actually,
the last component of (\ref{nv4}) does not arise from
the conservation law, since no such conservation law
exists for $P_T$. Rather it is a linearly dependent
equation that may be derived from the rest. However, we
include this time evolution equation for $v_T$ for
notational convenience.

As already described in section 2, the vectors $\vec n$ and
$\vec v$ are  defined as
\begin{equation}
\vec n \equiv \frac{\vec\pi}{ H},
\qquad
\vec v \equiv \frac{\vec P}{ H}.
\end{equation}
The two vectors, $\vec n$ and $\vec v$ satisfy the constraints
\begin{equation}\label{const0}
n^2 + v^2 =1, \qquad \vec n \cdot \vec v = 0.
\end{equation}
Thus, the dynamics produces a set of self-contained first-order
fluid equations for $H$ and $\vec n$ and $\vec v$. Although these
equations are more natural in Hamiltonian formulation, we stick
to Lagrangian formulation because the latter is more susceptible
to incorporation of curved background.

\subsection{Fluid Equations in Curved Space}

In the tachyon condensation limit ($V=0$),
the Hamiltonian in curved space is written by
\begin{equation}\label{curH}
{\cal H} = N \sqrt{\pi^m h_{mn} \pi^n
                  + P_m h^{mn}P_n} - L^m P_m,
\end{equation}
where we use $A_0=0$ gauge and the $X$ is defined in Eq.~(\ref{genX})
and to ensure the finiteness of Hamiltonian density we have to set
$X = 0$.
After some calculation, we obtain the following two relations
which is similar to the flat space case
\begin{eqnarray}
\frac{V}{\sqrt{\tilde X}} C^{mn}_A &=&
           = \frac{P^m\pi^n - \pi^m P^n}{T^{00}},
\nonumber \\
\frac{V}{\sqrt{\tilde X}}C^{mn}_S &=&
          = \frac{P^m P^n - \pi^m \pi^n}{T^{00}},
\label{ccof}
\end{eqnarray}
where $T^{00}$ is the (00)-component of the energy-momentum
tensor defined in Eq.~(\ref{emt}), and we define the upper
indexed $P^i$ as
\begin{equation} \label{upp}
P^m \equiv g^{m0} \pi^n F_{n0} + g^{mn}P_n.
\end{equation}
Using the relation in Eq.~(\ref{ccof}), we can rewrite the
equation of motion which is expressed in Eqs.~(\ref{geom}) as
follows~:
\begin{eqnarray}
& & \nabla_i\left(\frac{\pi^i}{\sqrt{-g}}\right) = 0,
\hskip 4.5cm  (\nu = 0\,\, \mbox{case}),
\nonumber \\
& & \nabla_0\left(\frac{\pi^i}{\sqrt{-g}}\right) + \nabla_j
\left(\frac{P^i\pi^j - P^j\pi^i}{\sqrt{-g} H}\right) = 0,
         \quad (\nu = i\,\, \mbox{case}),
\label{ceom}
\end{eqnarray}
where
\begin{equation}
H \equiv \sqrt{-g}\, T^{00} = \frac{1}{N\sqrt{h}}
                           \sqrt{\pi^m h_{mn}\pi^n + P_m h^{mn} P_n}.
\end{equation}
And the conservation of energy-momentum (\ref{cem})
is given by
\begin{eqnarray}
\nabla_0 \left(\frac{H}{\sqrt{-g}}\right)
           + \nabla_i \frac{P^i}{\sqrt{-g}} = 0 & &
        \qquad (\nu = 0\,\, \mbox{case}),
\nonumber \\
 \nabla_0\left(\frac{P^i}{\sqrt{-g}}\right) + \nabla_j
        \left(\frac{P^iP^j - \pi^i\pi^j}{\sqrt{-g}H}\right) = 0
         & &\qquad(\nu = i\,\, \mbox{case}).
\label{ccom}
\end{eqnarray}
This is an analog of (\ref{nv1}-\ref{nv4}) in curved spacetime.

The form of fluid equations take particularly simple form when
the shift vectors happen to vanish, $L^i =0$ and $N$ is constant.
This would be the case when we are considering a cosmological scenario,
for instance. With this, $\nabla_0$ effectively collapses to
$(1/N)\partial_t$,
and the Eqs.~(\ref{ceom}), (\ref{ccom}) are summarized by
\begin{eqnarray}
&&\vec\nabla' \cdot ( H \vec{\tilde n}) = 0,
\label{tnv1} \\
&& \dot H + \vec\nabla' \cdot ( H
              \vec {\tilde v}) = 0,
\label{tnv2} \\
&&\dot{\vec {\tilde n}} + (\vec {\tilde v}\cdot \vec\nabla')
               \vec {\tilde n}
               = (\vec {\tilde n} \cdot \vec\nabla') \vec {\tilde v},
\label{tnv3} \\
&&\dot{\vec {\tilde v}} + (\vec {\tilde v}\cdot \vec\nabla')
            \vec {\tilde v} =
            (\vec {\tilde n} \cdot \vec\nabla')\vec {\tilde n},
\label{tnv4}
\end{eqnarray}
where $\vec \nabla'=(\nabla_1',\nabla_2', 0)$ is the covariant derivative
defined for the spatial part of the metric, $h_{mn}$, and the dot
represent the time derivative $(1/N)\partial_t$.
We have also defined
\begin{equation}
\tilde n^m \equiv \frac{\pi^m}{H},
\qquad
\tilde v^m \equiv \frac{h^{mn} P_n}{H},
\end{equation}
with constraints
\begin{equation}
\tilde n^2 + \tilde v^2 = 1,
\qquad
\vec{\tilde n } \cdot \vec {\tilde v} = 0.
\end{equation}

\subsection{Integrability}

Since the quantities that enter the definition of ${\cal H}$,
$\vec n$, and $\vec v$, are naturally canonical variables,
solving the first-order fluid equation will not generate the elementary
fields $T$ and $A_\mu$ directly. For these, we have to solve for
another set of first order differential equations.

We could have derived the above fluid equation from the canonical
formulation, and if we did so, half of the canonical equations
would remain unused and simply relate conjugate momenta to
time-derivative of elementary fields. At least when we make use
of energy-momentum conservation directly. In terms of the above
variables, the remaining equations may be written as
\begin{equation}\label{half}
{\vec E} =  \vec n - \vec B \times \vec v.
\end{equation}
This determines the evolution of the ``magnetic field" $\vec B$
via Bianchi identity as
\begin{equation}\label{Bian1}
\dot{\vec B} = \vec\partial \times \vec n - \vec\partial\times(\vec B
               \times \vec v).
\end{equation}
Thus, $\vec B$ may be solved for after $\vec v$ and $\vec n$ are
determined. When this is done, $\vec E$ is also determined via (\ref{half}).

However, since $\vec B$ enters the fluid degrees of freedom via the
identity,
\begin{equation}
\vec v=\vec B\times \vec n,
\end{equation}
we need to make sure that this does not further restrict solutions to
the fluid equations alone. A priori, it is unclear whether for all
solutions to the fluid equations we can find $\vec B$ such that it
solves the evolution equation and is consistent with this algebraic
constraint as well. Only if such a $\vec B$ exist for the
solution to the fluid equation, we can say that the
solution is physical.

In the curved background these integrability conditions
are expressed as
\begin{eqnarray}
& & E_m = \frac1{N\sqrt{h}} \left( h_{mn} \tilde n^n
          + F_{mn}\tilde v^n\right) - F_{mn} L^n, \\
& & h_{mn}\tilde v^n = F_{mn} \tilde n^n.
\end{eqnarray}
plus the Bianchi identity
\begin{equation}
\dot F_{mn}=\partial_m E_n-\partial _m E_n.
\end{equation}

\section{Stability Analysis of Classical Solutions}

In this section, we perform stability analysis of ``static"
solutions of section 2. Because of lack of transverse pressure,
dimensionality of the unstable D-brane does not really
matter, so we will confine our analysis to the case of unstable
D2-brane. The authors of Ref.~\cite{Rey:2003zj} recently debated
against the assertion that the degeneracy
implies stability. Here we demonstrate the stability
explicitly by solving dynamical equations of motion,
and thereby invalidate the criticism.

Two cartesian coordinates are denoted as $x$ and $y$.
We will solve the Eq.~(\ref{nv1})-(\ref{nv4}) in flat space
using perturbation theory.
The Eqs.~(\ref{nv3}), (\ref{nv4}) may be conveniently rewritten as
\begin{equation}\label{ab1}
\dot{\vec a} - (\vec b \cdot\vec\partial)\vec a = 0,
\qquad
\dot{\vec b} + (\vec a \cdot\vec\partial)\vec b = 0,
\end{equation}
where we define
\begin{equation}\label{ab2}
\vec a \equiv \vec n + \vec v,
\qquad
\vec b \equiv \vec n - \vec v.
\end{equation}
The constraints (\ref{const0}) become
\begin{eqnarray}\label{const1}
a^2 = 1, \qquad b^2 = 1.
\label{const2}
\end{eqnarray}
Let us then consider the following small fluctuations around
the background fields $\vec a_0$, $\vec b_0$ which is given
by the ``static" classical solutions given in section 2. This
means that $\vec a_0 = \vec b_0 = \vec n_0$ with
\begin{equation}
{\vec n}_0 = (n_{0x}(y),0,n_{0T}(y)).
\end{equation}
Keep in mind here that  $0\le(n_{0T})^2=1-(n_{0x})^2\le 1$.

\subsection{First Order}

Expanding around such a solution
\begin{eqnarray}\label{expab}
\vec a &=& \vec a_0 + \vec a^{(1)}  + \vec a^{(2)}, \cdots,
\nonumber \\
\vec b &=& \vec b_0 + \vec b^{(1)}  + \vec b^{(2)}, \cdots,
\end{eqnarray}
and using the Eqs.~(\ref{ab1}), (\ref{expab}),
we obtain the first order perturbation equations
\begin{eqnarray}\label{fod1}
\partial_-\vec a^{(1)}
             &=& - \frac{\partial_y \vec n^{(0)}}{2 n_{0x}}b^{(1)}_y,
\nonumber \\
\partial_+\vec b^{(1)}
             &=& - \frac{\partial_y \vec n^{(0)}}{2 n_{0x}}a^{(1)}_y,
\end{eqnarray}
where we define
\begin{eqnarray}
x^{\pm} \equiv x \pm n_{0x} t, \qquad \partial_{\pm} =
\pm \frac1{2 n_{0x}} ( \partial_t \pm n_{0x} \partial_x).
\nonumber
\end{eqnarray}

\noindent
The $y$-components of the first order perturbation
equations in Eq.~(\ref{fod1}) are expressed by two homogeneous
first order differential equations
\begin{equation}\label{fod2}
\partial_- a^{(1)}_y = 0,
\qquad
\partial_+ b^{(1)}_y = 0,
\end{equation}
and the solutions are
\begin{equation}\label{fods1}
a^{(1)}_y = f(x^+,y), \qquad b^{(1)}_y = \tilde f(x^-,y),
\end{equation}
where $f(\tilde f)$ is an arbitrary function with arguments
$x^+(x^-)$ and $y$.

\noindent
Substituting the solutions for $a^{(1)}_y$ and $b^{(1)}_y$
into the $x$-components of the Eq.~({\ref{fod1}), we get
\begin{equation}\label{fod3}
\partial_- a^{(1)}_x
             = - \frac{\partial_y \vec n^{(0)}}{2 n_{0x}}
                 \tilde f(x^-,y),
\qquad
\partial_+ b^{(1)}_x
             = - \frac{\partial_y \vec n^{(0)}}{2 n_{0x}}
                 f(x^+,y),
\end{equation}
and the solutions are
\begin{equation}\label{fods2}
a^{(1)}_x = g(x^+,y) - \frac{n_{0x}'}{2 n_{0x}}
            \tilde F(x^-,y),
\qquad
b^{(1)}_x = \tilde g(x^-,y)
           - \frac{n_{0x}'}{2 n_{0x}} F(x^+,y),
\end{equation}
where $g$ and $\tilde g$ are arbitrary functions,
$n_{0x}'\equiv\partial_y n_{0x}$,  and we define
\begin{equation}
F(x^+,y) \equiv \int^{x^+} dw ~f(w,y),
\qquad
\tilde F(x^-,y) \equiv \int^{x^-} dw ~\tilde f(w,y).
\end{equation}
Using Eq.~(\ref{fod3}), we obtain free wave equations,
$\partial_+\partial_- a^{(1)}_x(b^{(1)}_x)=0$, i.e.,
\begin{equation}\label{fwav}
\left(\partial_t^2 - {n_{0x}}^2 \partial_x^2\right)n_x^{(1)}= 0,
\qquad
\left(\partial_t^2 - {n_{0x}}^2 \partial_x^2\right)v_x^{(1)}= 0,
\end{equation}
this means that there are propagating modes along background flux
line \cite{Gibbons:2002tv,Kim:2003he} and no net effects which deforms
the distribution of the background flux line and tachyon matter
in the first other perturbation.

\subsection{Second Order: Low Frequency Limit}

Now let us consider the second order perturbation
of the Eq.~(\ref{ab1}). General formalisms for the second order
perturbation are analyzed in Appendix C.
As a meaningful choice for the solution of the first order
perturbation equations to investigate the second order ones,
we consider the following configurations
\begin{eqnarray}
f(x^+, y) &=& u(y) + c(x^+,y), \\
\tilde f(x^-, y) &=& -u(y) + \tilde c(x^-,y), \\
g(x^+,y) &=& - \frac{n_{0x}'}{2 n_{0x}} u(y) x^+ + d(x^+,y), \\
\tilde g(x^-,y) &=&  \frac{n_{0x}'}{2 n_{0x}} u(y) x^-
                     + \tilde d(x^-,y),
\end{eqnarray}
where $c(\tilde c)$ and $d (\tilde d)$ are arbitrary oscillatory functions.
In other words, we allow a net velocity $u(y)$ along $y$-direction
in the first order perturbation.

Since the second order perturbation is pretty involved, let us take
some simplifying limit.
We could for instance consider taking a very low frequency
limit. After all, if there is a confining or dispersive effect, it should
show up here. In the current general setup, we may achieve this by taking
the limit
\begin{equation}
f_{osc}^{(m,n)}(x^{\pm},y) \rightarrow 0,
\end{equation}
where $f_{osc}$ represent oscillatory piece and $m$, $n$ are arbitrary
positive integers. We use the notation,
\begin{equation}
A^{(m,n)}(x,y) \equiv \frac
{\partial^{m+n}}{\partial x^m \partial y^n} A(x,y).
\end{equation}
In addition, we average over characteristic time-scale of the oscillatory
pieces, meaning that we are mainly interested in net effect rather than
exact time evolution of the system. We achieve this by averaging the equation
over time where in effect we set
\begin{equation}
\langle f_{osc} \rangle \equiv \lim_{L \to \infty} \frac1{L}
     \int^{\frac{L}2}_{-\frac{L}2} dw \, f_{osc}(w,y)
      \rightarrow 0.
\end{equation}
In this limit, the Eqs.~(\ref{sod3}) - (\ref{sod6}) are reduced to
\begin{eqnarray}
& & (\partial_t^2 - {n_{0x}}^2 \partial_x^2) n^{(2)}_x
                     = 2 uu' n_{0x}' + u^2 n_{0x}'',
\label{n2}\\
& & (\partial_t^2 - {n_{0x}}^2 \partial_x^2) v^{(2)}_x = 0,
\\
& & (\partial_t^2 - {n_{0x}}^2 \partial_x^2) n^{(2)}_y = 0,
\\
& & (\partial_t^2 - {n_{0x}}^2 \partial_x^2) v^{(2)}_y = 0.
\end{eqnarray}
Left-hand-sides has the Klein-Gordon kinetic operator for an
one-dimensional system with ``speed of light" equal to
$\sqrt{1-n_{0T}^2}=|n_{0x}|$. This is identical to the first order
case, and implies that small fluctuation would move freely
up and down along the flux lines, provided that  there is no
term on the right hand side.

On average, there is no net
force on the second order fluctuation, generated from the
first order fluctuation, except the right hand side of (\ref{n2}).
However, this
term represents a rather trivial effect. $u(y)\neq 0$ implies that
the entire configuration is drifting along $y$ direction. Since
the flux lines and tachyon matter are distributed nontrivially
along $y$ direction, this motion will generate time-dependent
change of $\vec n$. To see what effect it has, we need to solve for
$v^{(2)}$ induced by $u(y)$ first. For $x$-independent perturbation,
we have
\begin{equation}
\dot{\vec v}+(\vec v\cdot \vec \partial) \vec v =0,
\end{equation}
which gives
\begin{equation}
v_y^{(1)}(y,t)+v_y^{(2)}(y,t)=u(y)-u(y)u'(y)t.
\end{equation}
With this velocity field, we may ask
how $\vec n(y,t)$ drifts with time. If the only physical effect is the
drift, the $\vec n(y,t)$ will be identical to
$\vec n(\tilde y,0)=\vec n(y)$ where
\begin{equation}
y=\tilde y +\int_0^t  ds \;v_y(f(s),s) ,
\end{equation}
with $f(s)$ is the $y$-trajectory between $f(0)=\tilde y$ and $f(t)=y$ due
to the velocity field $v_y(y,t)$. Despite somewhat involved formulae so
far, the relationship between $y$ and $\tilde y$ is deceptively simple,
\begin{equation}
y=\tilde y + u(\tilde y) t +O(t^3),
\end{equation}
or equivalently
\begin{equation}
\tilde y =y- u(y) t + u(y)u'(y) t^2 +O(t^3).
\end{equation}
Thus we find simple drift of the configuration along $y$-direction
gives,
\begin{eqnarray}
&&\vec n(y,t)=\vec n(\tilde y)\nonumber \\
&=&\vec n(y) -t\left(u(y)\partial_y\vec n(y)\right)
+\frac{t^2}{2}\left(2 u(y)u'(y)\partial_y \vec n(y)
+u(y)^2\partial_y^2\vec n(y)\right) , \label{drift}
\end{eqnarray}
up to 2nd order in time $t$. Thus the nontrivial term on the
right hand side of (\ref{n2}) simply represent this drift effect.\footnote{
We could have started perturbation after suppressing oscillatory
pieces completely, instead of averaging over it, to see such drift
effect. If we did that, we would have found
exactly (\ref{drift}) as the solution.}

\subsection{Second Order: Case of Interlocked Distribution of the Two Fluid}

One large subset of classical solutions we could consider in more
detail is those with
$n_{0x}$ constant. Since $n_{0x}$ is the ratio between the flux
energy density and the total energy density, such a solution
corresponds to an arbitrary distribution of the energy density
along $y$ direction, while maintaining the ratio $\pi_{0x}/\pi_{0T}$
fixed. In this case, all terms with derivative on $n_{0x}$ die away, so
we have
\begin{eqnarray}
\partial_+ \partial_- a^{(2)}_x
                &=&
                 -  \frac1{2 n_{0x}}\tilde f g^{(1,1)}
                 - \frac1{2 n_{0x}} \tilde g g^{(2,0)},
\label{sod3'} \\
\partial_+ \partial_- b^{(2)}_x
                &=&
                  -  \frac1{2 n_{0x}} f \tilde g^{(1,1)}
                   - \frac1{2 n_{0x}}  g
                  \tilde g^{(2,0)},
\label{sod4'} \\
\partial_+\partial_- a_y^{(2)}
           &=&
           -\frac{1}{2 n_{0x}} \tilde f f^{(1,1)}
            -\frac{1}{2 n_{0x}}\tilde g f^{(2,0)},
 \label{sod5'} \\
\partial_+\partial_- b_y^{(2)}
          &=&
           -\frac{1}{2 n_{0x}}  f \tilde f^{(1,1)}
           -\frac{1}{2 n_{0x}} g \tilde f^{(2,0)}.
\label{sod6'}
\end{eqnarray}
Since we saw the effect of $u(y)$ is an overall drift of the
system along $y$ direction, we could safely turn it off in the
first order perturbation, $f$, $g$, $\tilde f$, and $\tilde g$.
Then the first two are oscillatory function of $x^+$ while the
latter two are oscillatory functions of $x^-$. All terms
on the right hand side are of the form,
\begin{equation}
h(x^+)\times\; \tilde h(x^-) ,
\end{equation}
for a pair of some oscillatory functions $h$ and $\tilde h$.
Such combination of force terms on the right hand side cannot
generate a net effect, when averaged over time, since the
resonance effect cannot occur. The right hand sides will
drive some oscillation of $a$ and $b$ and do not lead to
any instability.

Thus, we conclude that this large class of classical solutions
are all stable under perturbation and any possible fluctuations
move freely along a direction set by electric flux lines. The
origin of this one-dimensional behavior was previously explained
in terms of collapse of an effective-causal-structure. We again
emphasize that the integrability condition can at most restrict
acceptable solution to the fluid equations, so stability under
the latter is sufficient to argue the stability under the
full field equation.

\section{Hamiltonian Dynamics with Potential $V$}

Much of what we studied above concern a region where tachyon rolls
to one side of potential. For example we are imagining that
$T\rightarrow \infty$ everywhere. In this section, we take into
account possibility of $V\neq 0$ in the dynamics. The formulation
here should serve useful tool for understanding initial
stage of tachyon condensation.

\begin{figure}
\centerline{\epsfig{figure=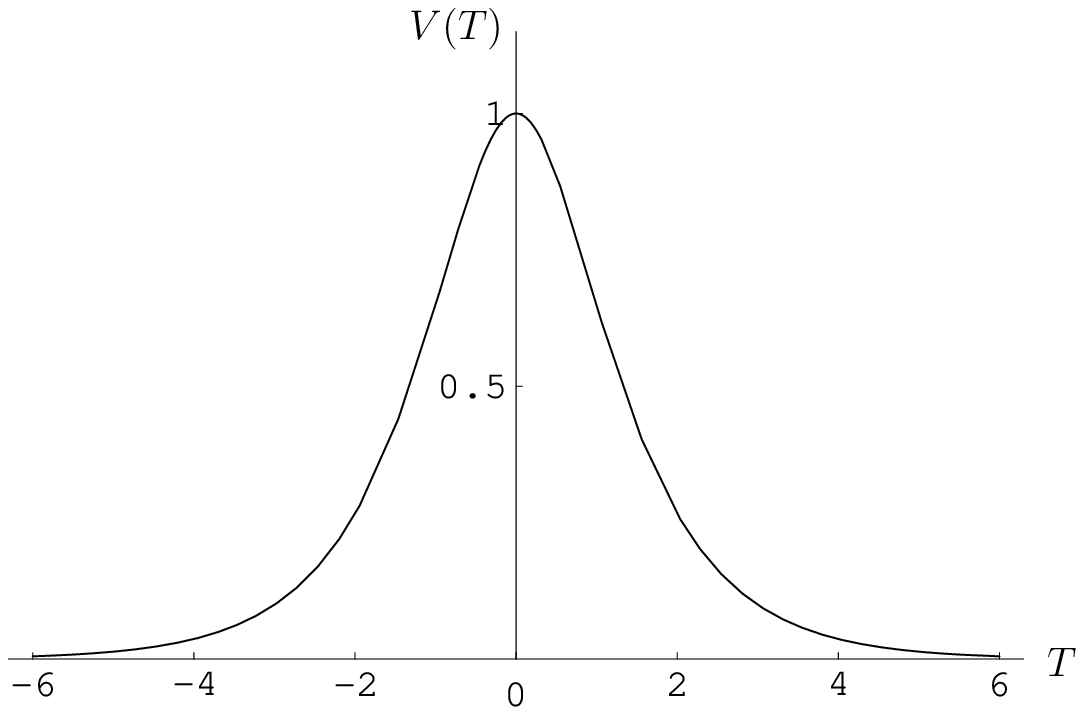,height=6cm}}
\begin{quote}
\caption{A prototypical form of the potential $V(T)$ as function
of $T$, $V(T)=1/{\rm cosh}T$ \cite{Kutasov:2003er,Kim:2003he,
Buchel:2002tj, Leblond:2003db,Lambert:2003zr}. We will use this potential
for plot of the time evolution of flux later in this section.}
\end{quote}
\label{fig1}
\end{figure}

\subsection{Canonical Field Equations with $V$}

Let us consider the dynamics with $V\neq 0$ somewhere.
For simplicity, we will consider unstable D2 brane
case again, with all transverse scalars suppressed.
Recall that we are using the notation introduced in section 2,
where a fictitious direction $x^T$ is employed and  $T$ is
treated as if it is a component of the
gauge field along $x^T$. For more details we
refers the reader to  Eqs.~(\ref{EB}), (\ref{fmom}).

Half of the Hamiltonian equations of motion
\begin{eqnarray}
& & \vec E = \frac{1}{{H}} ( \vec \pi - \vec B \times \vec P),
\label{nve1}
\end{eqnarray}
generates the evolution equation for
$\vec B =(\partial_y T,-\partial_x T, B)$ when combined with
the Bianchi identity
\begin{equation}
\dot{\vec B}
=\vec\partial \times \vec E.
\end{equation}
It is important to note that the energy density $H$
has a $V^2$ term inside the square root,
\begin{equation}\label{nvh}
{H} = \sqrt{\vec \pi^2 + \vec P^2 + V^2 (1+\vec B^2)}.
\end{equation}
The other half gives evolution equations for electric flux
$\pi_{x,y}$ and conjugate momentum for tachyon $\pi_T$,
\begin{eqnarray}
& & \dot \pi_i + \partial_j \left(\frac{\pi_i P_j - \pi_j P_i
           + V^2 F_{ij}}{{H}} \right) = 0,
\label{nve2}\\
& & \dot \pi_T + \partial_j \left(\frac{\pi_T P_j - \pi_j P_T
           - V^2 \partial_j T}{{H}} \right)
           = - \frac{VV'(1 + {\vec B}^2)}{{H}}
\label{nve3}
\end{eqnarray}
with $V' \equiv {\partial V}/{\partial T}$ and $F_{ij}=B \epsilon_{ij}$.

These evolution equations should be consistent with energy-momentum
conservation which now takes the modified form
\begin{eqnarray}
& & \dot {H} + \partial_i P_i = 0,
\label{nvc1} \\
& & \dot P_i + \partial_j\left( \frac{P^i P^j -
        \pi^i \pi^j - V^2(\delta^{ij} + B^i B^j)}{{H}}\right) = 0,
\label{nvc2}
\end{eqnarray}
which is almost identical to the previous case except for $V^2$
terms. We use the following facts which are the flat metric
version of
Eq.~(\ref{emt}),
\begin{eqnarray}
T^{00} &=& \frac{V}{\sqrt{\tilde X}}C^{00}_S= {H},
\nonumber \\
T^{0i} &=& \frac{V}{\sqrt{\tilde X}}C^{0i}_S = P^i,
\nonumber \\
T^{ik} &=& \frac{V}{\sqrt{\tilde X}} C^{ik}_S = \frac{P^i P^k -
        \pi^i \pi^k - V^2(\delta^{ik} + B^i B^k)}{{H}}.
\nonumber
\end{eqnarray}
In all of above equations, it is quite clear that the fluid-like behavior
of $V=0$ regime will be ruined by the potential, and its effect is
of order $V^2$.

\subsection{Flux Motion near Domain Wall Formation}

In tachyon condensation, $V\neq
0$ can survive the decay process if there is a topological defect.
In case of single unstable D$p$ branes, only possible topological
defect would be  domain walls, separating a region of $T=\infty$
from that of $T=-\infty$, and at the end of day become  $D(p-1)$
branes in the context of type II theories.
We will concentrate on
initial configuration which will lead to a single flat D$(p-1)$
brane. With such initial configuration, we would like to ask
how flux behaves where $V \ne 0$.

To answer this question, we consider an initial configuration
of $T$ such that it has spatial variation along $y$ direction
and vanishes at $y=0$. In addition we assume static
initial condition, so that neither the tachyon matter nor
the string fluid has initial velocity. Finally for the
sake of simplicity, we further assume a uniform distribution
of flux lines, lined up along $x$ direction. This can be
summarized by the following initial conditions,
\begin{equation} \label{init}
T_0 = T(y) , \qquad \vec \pi_0 = (\pi_{0x}, 0, 0), \qquad B=0,
\end{equation}
where $\pi_{0x}$ is a constant.

\begin{figure}
\centerline{\epsfig{figure=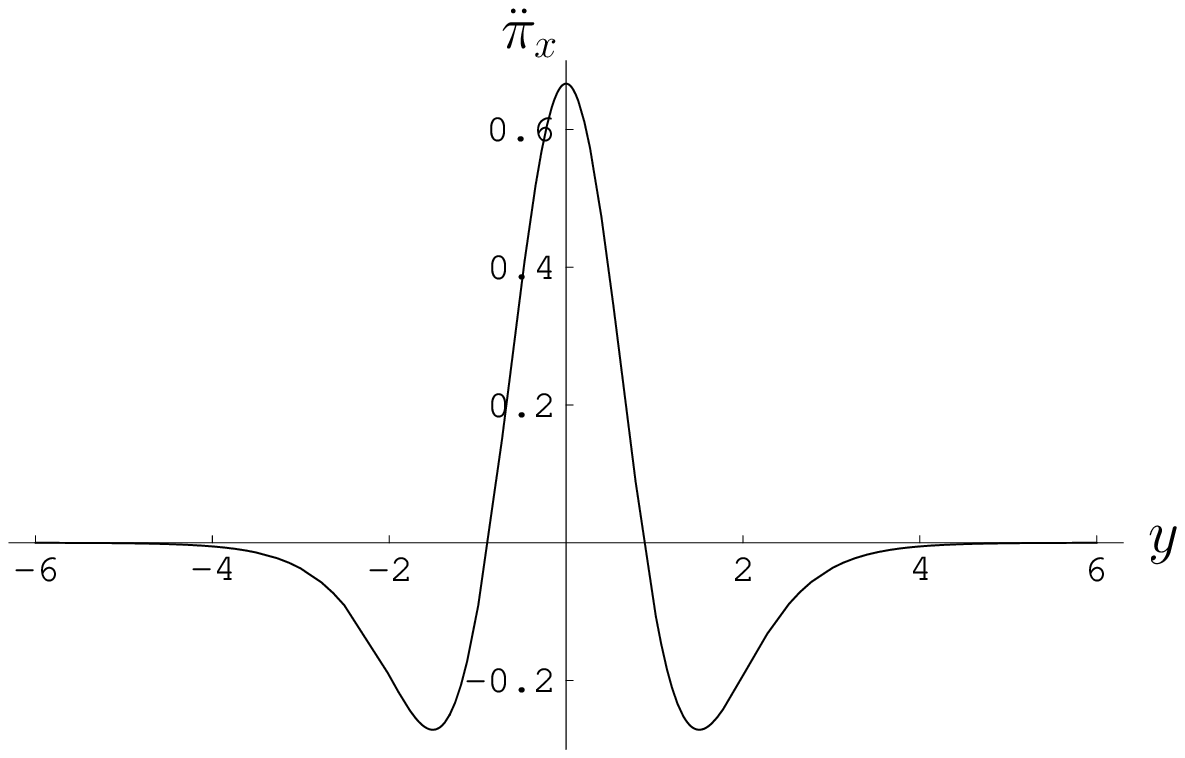,height=6cm}}
\begin{quote}
\caption{Plot of $\ddot\pi_x$ as function of $y$, with
$V(T)=1/{\rm cosh}T$ and $T=y$. It shows short range
attraction of fluxes toward $T=0$ during domain wall
formation. At late time, $T=uy$ with $u\rightarrow \infty$,
so the range of the attractive force infinitesimal. The
final configuration at $u=\infty$ corresponds to
BPS D$(p-1)$ brane with some fundamental string flux trapped.}
\end{quote}
\label{fig2}
\end{figure}

\begin{figure}
\centerline{\epsfig{figure=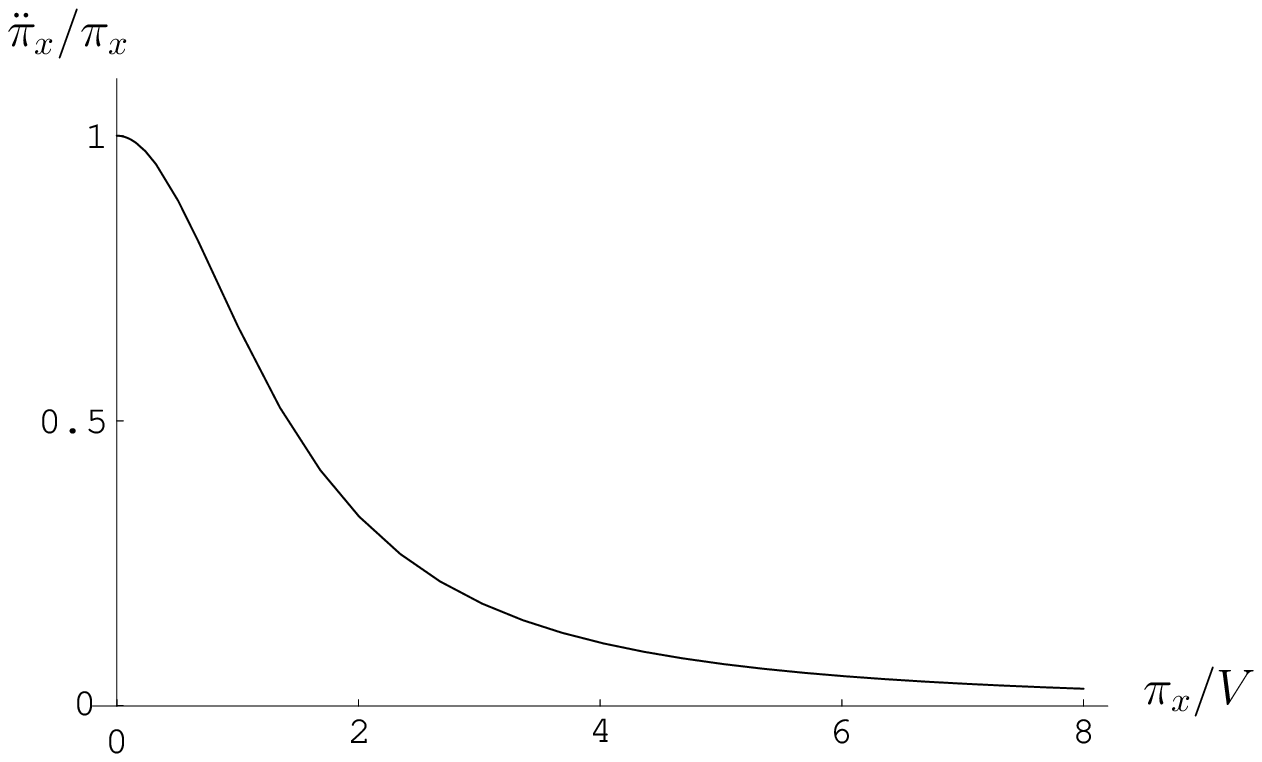,height=6cm}}
\begin{quote}
\caption{ Plot of
$\ddot\pi_x/\pi_x$ at $y=0$ as a function of $\pi_x/V$ at $y=0$,
again at $t=0$ with the initial condition $T=y$.}
\end{quote}
\label{fig3}
\end{figure}

Now let us consider evolution of $\pi_i$, ($i=x,y$)  as follows,
\begin{equation}\label{tayl}
\pi_i(t) = \pi_{i0} + t \dot \pi_i|_{t=0}
            + \frac12 t^2 \ddot \pi_i |_{t=0} + \cdots.
\end{equation}
Under the initial condition (\ref{init}), the first non-trivial
variation of the flux appears in $\ddot \pi_i |_{t=0}$ term
in Eq.~(\ref{tayl}), i.e., $\dot \pi_i|_{t=0}=0$, and is given
by\footnote{Note that at $t=0$ the energy density $H$ reduces to
\begin{eqnarray}
\sqrt{\pi_x^2 + V^2\left(1 + (\partial_y T)^2\right)},
\nonumber
\end{eqnarray}
while
\begin{equation}
\dot B = -\partial_y \left(\frac{\pi_x}{{H}}\right)_{t=0},
\qquad
\dot P_x = 0,
\qquad
\dot P_y = \partial_y \left(\frac{V^2}{{H}}\right)_{t=0}.
\end{equation}}

\begin{eqnarray}
\ddot \pi_x &=& - \partial_y\left(\frac{\pi_x \dot P_y
                  + V^2 \dot B}{{H}}\right)_{t=0}
\nonumber \\
           &=&  \left[\frac{V^2}{{H}} \partial_y^2
               \left(\frac{\pi_x}{{H}}\right)
               -\frac{\pi_x}{{H}} \partial_y^2
               \left(\frac{V^2}{{H}}\right)\right]_{t=0},
\label{nve4}
\end{eqnarray}
while $\pi_y$ is not generated up to this order,
\begin{eqnarray}
\dot \pi_y=0,\qquad \ddot \pi_y =0.\label{nve5}
\end{eqnarray}
Thus, we find the leading order time-variation of flux lines is summarized
as
\begin{equation}
\pi_x=\pi_x^0+\frac12 t^2 \left[\frac{V^2}{{H}} \partial_y^2
               \left(\frac{\pi_x}{{H}}\right)
               -\frac{\pi_x}{{H}} \partial_y^2
               \left(\frac{V^2}{{H}}\right)\right]_{t=0} + O(t^3).
\end{equation}
{}From  this expression, it is not too difficult
to show that initially uniform $\pi_x$ will get redistributed such that
flux in the region $V\sim 1$ tends to gather toward $T=0$. Since this effect
is of order $V^2$, which is exponentially small far away from $T=0$ domain
walls, the interaction is of short range. The effective range of
this attraction is determined by values of $V$ and therefore by
the gradient of $T$ along $y$. With $T=uy$ and in unit where
$\alpha'\sim 1$, the range is roughly $\delta y\sim 1/u$. Thus, the
effect is maximal when the tachyon begins to roll and die away quickly
as the tachyon condensation progresses; To reach final stationary state,
$u$ has to be infinite, as is well known.

To illustrate this effect graphically, we draw two figures.
Figure 2 is the plot of $\ddot\pi_x$ as function of $y$,
assuming an initially linear gradient of $T=y$. This clearly
shows local attraction of fluxes toward $T=0$. Final figure is a plot of
$\ddot\pi_x/\pi_x$ at $y=0$ as a function of $\pi_x/V$ at $y=0$, which
shows that the attraction is universal and tends to get weakened when
flux energy dominates over the tachyon energy.
This effect explains the phenomenon observed in  Ref.~\cite{Rey:2003zj}
and produces D$(p-1)$ branes and fundamental strings.

\subsection{Confined Flux Strings are Exceptional and Rare Final State}

Some of more recent proposals for fundamental string formation utilizes
configurations involving $T=0$, which would be center of domain wall
separating regions of $T\rightarrow \infty $ and of $T\rightarrow -\infty$
\cite{Hashimoto:2002sk, Rey:2003zj,Hashimoto:2003qx}.
In the above, we analyzed the classical dynamics in such situations, and gave a
quantitative description on aggregation of flux lines near domain walls
with $T=0$ at the center. We should
note here that any such process will involve bound states of fundamental
string with $D(p-1)$ branes as the final product, and the fundamental
string flux will be confined to a co-dimension-one hyper-surface rather
than a thin, string-like, flux bundle.

We must emphasize here that the real question lies not in whether there
are confined flux string but rather how such confined configuration become
generic configuration. In fact, classical solution (even with $V=0$
everywhere) that behaves exactly like a Nambu-Goto string has been
introduced in Ref.~\cite{Gibbons:2000hf} and further studied in
Ref.~\cite{Sen:2000kd}. This classical solution admits infinitely
many marginal deformation that disperse the flux lines, which makes
it very unlikely configuration to form, to begin with.

Decay processes involving
topological defects are intriguing in that it could
lead to this Nambu-Goto string by fine-tuning initial configurations.
For instance, one could imagine a infinitely long cylinder,
$R\times S^{p-2}$, of a domain wall to which nearby electric flux
get attached. Depending one details of the initial condition, it is
possible to imagine that $S^{p-2}$ part of the cylindrical domain wall
collapses after tachyon condensation completes and
perhaps push all electric flux on its world-volume
to an infinitesimally thin string. Such configuration will also
obey a Nambu-Goto dynamics classically \cite{Sen:2003bc}.
However, the trouble is simply that, in the phase space of
this theory, such string-like configurations are very exceptional
ones. One must still understand exactly how this huge degeneracy
is lifted, be it from higher derivative correction or via
some quantum effect \cite{Yi:1999hd,Bergman:2000xf,Kleban:2000pf}.

\section{BPS Limit and Strings Orthogonal to D$(p-1)$ Branes}

Fundamental strings orthogonal to a D$(p-1)$ brane 
preserves 1/2 of supersymmetry left unbroken by the D-brane, so
the system of such fundamental strings and a D-brane should
preserves 1/4 of 32 supersymmetry in spacetime. With the
domain wall incorporated into the system, we could ask whether 
there is such a BPS-like solution involving string fluid
transverse to D$(p-1)$ branes. Generic solutions with right
spacetime charges was written down in Ref.~\cite{Kim:2003in}.

This smoothness of the domain wall solution may be contrasted to the
singular solutions  considered in Ref.~\cite{Sen:2003tm}. This smoothness
of the solution caution us against in using singular
domain walls in the presence of transverse string fluid. It was
recently argued that string fluid would be repelled by D$(p-1)$
brane, if the former try to impinge upon the latter \cite{Sen:2003bc}.
This is based on energetics of the string fluid assuming a singular 
domain wall background. The smoothness of the solution above is generically
attributable to the presence of transverse string fluid, which shows
that backreaction due to string fluid may be  important in some cases.

In this section, we will rediscover these smooth solution via our
formulation. In particular, we find that a minimal solution of a
sort exists and saturate the right BPS energy bound, expected of
fundamental strings orthogonal to a D$(p-1)$ brane.

\subsection{Smooth Domain Walls Threaded by String Fluid}

When $V$ is included in the analysis, we may consider a different
kind of static solutions, involving domain walls. A singular
solution of domain wall type has been known \cite{Sen:2003tm},
but recently it was found that a smooth domain wall solution
is possible when we modify the asymptotic boundary condition
\cite{Kim:2003in,0305092,Brax:2003rs}.
Here we will reproduce these solutions within our formalism
and study its properties.

For this, we
will consider static configurations with $P_i=0$ and $\pi_T=0$, and
assume a uniform electric flux, $\pi_x$. We will achieve $P_i=0$ by
setting $F_{ij}=0$. With this, another allowed quantity is
\begin{equation}
v_T=n_x\partial_x T.
\end{equation}
In such configurations, a useful identity is
\begin{equation}
H=\sqrt{(\pi_x^2+V(T)^2)(1+(\partial_x T)^2)}.
\end{equation}
Then the momentum conservation (\ref{nvc2}) forces
\begin{equation}
0=\partial_x\left(\frac{\pi_x^2+V(T)^2}{H}\right)
=\partial_x\left(\frac{\pi_x^2+V(T)^2}{1+(\partial_x T)^2)}\right)^{1/2},
\end{equation}
which may be rewritten as
\begin{equation}
C^2\left(\pi_x^2+V(T)^2\right)=1+(\partial_xT)^2
\end{equation}
for some constant $C$. Inspection of (\ref{nve1}) shows that
this constant is related to the asymptotic value of $E_x\rightarrow
\epsilon$ as $(C\pi_x)^2=\epsilon^2$.\footnote{For the present
solutions $E_x$ itself is a constant, by the way.} Thus we only
need to solve
\begin{equation}
\frac{dT}{dx}=\pm \sqrt{\epsilon^2(1+V(T)^2/\pi_x^2)-1}.
\end{equation}
There are three generic cases to consider
\begin{itemize}
\item
If $\epsilon^2$ is larger than or equal to 1, 
this gives monotonic solution $T(x)$,
and represents a single D$(p-1)$ (or anti-D$(p-1)$) brane at $T=0$.

\item
If $\epsilon^2$ is less than 1 but larger than $1/(1+V(0)^2/\pi_x^2)$,
this gives an oscillatory $T(x)$ with an infinite array of zeros and
represents an alternating array of D$(p-1)$ and anti-D$(p-1)$ branes.

\item
If $\epsilon^2$ is less than  $1/(1+V(0)^2/\pi_x^2)$, no solution exists.

\end{itemize}
In all cases where solutions exist, the configuration is that of smooth
domain wall solutions, generically threaded by uniform distribution of
flux lines transverse to the domain wall.

\subsection{A Minimal Solution and BPS Energy}

Energy density of the above solutions are
\begin{equation}
H=\sqrt{(\pi_x^2+V^2)(1+(\partial_xT)^2)}
=\frac{\epsilon}{|\pi_x|}(\pi_x^2+V^2)
\end{equation}
which approaches $\epsilon\pi_x$ asymptotically. 
We are interested in solution with pure string 
fluid far away from domain wall, and this forces us to consider 
$\epsilon=1$, in particular. This choice is also 
energetically favored since it represents the lowest energy 
solution available, given the conserved flux and  the single 
domain wall. As we will see shortly, this case may be regarded 
as a BPS-saturated solution.

As would be expected from $E_x^2=1$, the equation of
motion simplifies to
\begin{equation}
\frac{dT}{dx}=\pm \frac{V(T)}{\pi_x}
\end{equation}
whose solution is such that $T$ does approach
vacuum at $x=\pm \infty$. The energy density also simplifies further 
for $\epsilon=1$ as\footnote{We are indebted to Chanju Kim on this point.},
\begin{equation}
H=|\pi_x|+\frac{V^2}{|\pi_x|}=\pi_x+ V|\partial_x T|
\end{equation}
The first piece is the BPS energy density associated 
with the fundamental string charge. The second piece must be
associated with the domain wall itself. In fact, integrating
over $x$, this is precisely the tension of the D$(p-1)$ brane
\begin{equation}
\int V\partial_xT\,dx=\int_{-\infty}^\infty VdT =\tau_{p-1}
\end{equation}
The energy
of the solution is then precisely sum of two terms, one from tension
of the D$(p-1)$ brane and the other from transverse fundamental string.
This gives exactly the BPS energy expected of fundamental string ending on
or passing through a D-brane.

Thus at least when we have uniform string fluid threading a 
D$(p-1)$-brane, the string fluid can mimic, quantitatively,
behavior of fundamental string ending on D-branes. This may be
compared with findings of Ref.~\cite{Sen:2003bc}, where it is asserted
that string fluid cannot end on D$(p-1)$ brane. The latter is based 
on a computation where an infinitely thin D$(p-1)$ is taken to be a 
background and neglects a possible backreaction of the domain wall 
solution to the presence of string fluid. While the assertion might
stand when transverse string flux are well-isolated, this example
cautions us to be careful about backreaction of a domain wall.

\section{Summary }

We have reviewed dynamics of the tachyon coupled to a gauge field.
When the tachyon condensation has progressed far so that $V\rightarrow
0$, the dynamics is that of two fluids, string fluid and tachyon
matter, as anticipated, but the fluid equation of motion must be
augmented by a set of integrability condition, which is necessary
if we wish to recover $A_\mu$ and $T$ from the fluid variables.
We have isolated a large family of static solutions with huge degeneracy
and tested their stability. 

We further extended the formalism to
the case of $V\neq 0$ somewhere, and showed that during initial
stage of condensation, electric flux lines tends to be attracted
toward $T=0$. Since $T=0$ may survive the condensation process
only if there is a domain wall, this initial configuration will
lead to a D$(p-1)$ brane with fundamental string flux spread on it.
Interaction between a domain wall and transverse string fluid is
more drastic, and the latter thickens the former. We also wrote
down such smooth domain wall solution, using the Hamiltonian
formulation, and discovered that minimal solution of this type 
saturates a BPS bound of strings ending on D-branes.

One aspect of the $V=0$ limit, worth emphasizing, concerns the coupling
between the two fluid components. Despite a
rather tight coupling between the string fluid and the tachyon matter,
as evidenced by the form of the Hamiltonian, static distribution of the
two fluid components seems pretty much independent of each other. The only real
constraint is that static distribution of tachyon matter
must be uniform along the string fluid direction. One might have
expected that two fluid components come with the same (transverse)
distribution, but this is not the case at all.
Instead, the tight coupling between the tachyon matter and the string
fluid affects the subsequent dynamics in an unexpected way.  As is clear
from the fluid equation, distribution of string fluid and distribution
of tachyon matter evolves in the complete absence of pressure (except
for the tension along the string fluid) and is essentially free.
That is they simply responds to velocity field of fluid, which is
in turn affected by gradient of density of flux direction along the
flux lines. Regardless of the details of the latter, however, two
fluid shares one and the same velocity field $v$. This implies that
once the relative distribution of the two fluid component is set, the
subsequent perturbation moves string fluid and tachyon together
in such a manner that keeps this ratio fixed when followed by co-moving
observers.

The classical system of tachyon and gauge field turned out to have many
intriguing surprises; fluid-like behavior of the perfectly sensible field 
theory, the drastic reduction of perturbative degrees of freedom, Carollian
collapse of effective light-cone, and finally many intriguing classical 
solutions, static or dynamic, singular or nonsingular. One question that 
remains murky is whether this classical field theory admits a sensible 
and practical quantum treatment of its own, regardless of its tie to
stringy context, and whether interesting physics arise from a quantization. 
We feel that this avenue of research deserves more study in future.

\vskip 1cm

\leftline{\large \bf Acknowledgments}
\vskip 0.5cm
\noindent
We thank
Chanju Kim, Yoonbai Kim, and Kimyeong Lee for useful conversations.
O.K. is supported by Korea Science $\&$ Engineering
Foundation through Astrophysical Research Center for the
Structure and Evolution of the Cosmos (ARCSEC). P.Y. is supported 
in part by Korea Research Foundation (KRF) Grant KRF-2002-070-C00022.

\appendix

\section{S-Dual Formulation of $V=0$ Limit}

Ref.~\cite{Gibbons:2000hf} discovered a nonvanishing dual
Lagrangian for pure gauge system, whose equation of motions and Bianchi
identity are respectively
Bianchi identity and equation motion of the original Born-Infeld
system. It is rather clear that dualization procedure works even when
scalar fields are present, once we think of the latter as components of
the gauge field along some fictitious dimensions.

The prescription for the dualization process is to introduce a new
gauge field $C$ and its field strength, $G=dC$ such that
\begin{equation}
{\cal K} \equiv \pi_m\,dt\wedge dx^m+ \frac12
\,K_{mn}dx^m\wedge dx^n=
*G.\label{hodge}
\end{equation}
Momentum $\pi_m$ is canonical conjugate momenta of $A_m=(A_1,\cdots ,
A_p,  T)$. $K_{mn}$ are defined as
\begin{equation}
K_{mn}=\frac{\delta {\cal H}}{\delta F_{mn}}.
\end{equation}
Finally, the Hodge-dual operator $*$ here is defined with respect to
the ``volume form"
\begin{equation}
dt\wedge dx^1\wedge dx^2\wedge \cdots\wedge dx^p\wedge dx^T.
\end{equation}
With this, the Legendre transformation of the Hamiltonian gives a
dual Lagrangian as a function of the dual field strength $G$,
\begin{equation}
{\cal L}'=\sqrt{-{\cal K}^2/2}=\sqrt{G^2/2}.
\end{equation}
One crucial point here is that the Legendre transformation that
leads to ${\cal L}'$ is not reversible. Thus, to obtain truly
dual Lagrangian, one must introduce constraints on the dual side.

As explained in Ref.~\cite{Gibbons:2000hf}, this does not complete
the dualization process
due to the fact that the original Hamiltonian is quite degenerate.
That is,  part of $F_{mn}$ which has no inner product with $\pi_m$
never enters the Hamiltonian. Because of this, inverse Laplace
transform of ${\cal L}'$ does not give us back $\cal H$. This
problem is curable by adding constraints,
\begin{equation}
{\cal K}\wedge {\cal K}=0.
\end{equation}
Thus the correct dual Lagrangian may be written as
\begin{equation}
\sqrt{G^2/2} +\langle \lambda, *G\wedge *G\rangle
\end{equation}
with a Lagrange multiplier field $\lambda$. This is where the
pure gauge dynamics maps to that of Ref.~\cite{Nielsen}, which
attempted to write a field theory that produces string-like
degrees of freedom classically.

\section{Canonical Field Equation with $V=0$}

The tachyon effective action with gauge field is given by
\begin{equation}\label{gta}
S = \int d^{p+1} x \ {\cal L},
\end{equation}
where
\begin{eqnarray}
{\cal L} &=& - V(T) \sqrt{- X},
\nonumber \\
X_{\mu\nu} &\equiv& g_{\mu\nu} + F_{\mu\nu}
                    + \partial_\mu T \partial_\nu T,
\nonumber \\
X &\equiv&  \det X_{\mu\nu}.
\end{eqnarray}
Then the determinant $X$ can be expressed as
\begin{equation}\label{detX2}
X = X_{00} \det (X_{ij}) - X_{0i} D^{ij} X_{j0},
\end{equation}
where the matrix $D$ is transpose of the cofactor for matrix $(X)_{ij}$ and
the components of matrix $(X)_{\mu\nu}$ are written by
\begin{eqnarray}
X_{00} &=& - N^2 + h_{ij}L^i L^j + \dot T^2, \nonumber \\
X_{0i} &=&  E^+_i, \nonumber \\
X_{i0} &=& - E^-_i, \nonumber \\
X_{ij} &=& h_{ij} + F_{ij} + \partial_i T\partial_j T\nonumber
\end{eqnarray}
with
\begin{equation}
E_i^{\pm} \equiv F_{0i} \pm \dot T \partial_i T \pm L^j h_{ji}.
\end{equation}

Let us denote the Hamiltonian using canonical variables to describe
the dynamics of the system by Hamiltonian equations.
Then conjugate momenta are given by
\begin{eqnarray}
\pi^i &\equiv& \frac{\partial {\cal L}}{\partial \dot A_i}
               = \frac{V}{\sqrt{-X}} \frac{E^+_j D^{ji}
                 + D^{ij} E^-_j}{2},
\label{ggmom} \\
\pi_T &\equiv& \frac{\partial {\cal L}}{\partial \dot T}
           = \frac{V}{\sqrt{-X}}
             \left(\dot T \det X_{ij}
              - \frac{E^+_j D^{ji}\partial_i T
              -\partial_i T D^{ij}E^-_j}{2}\right),
\nonumber \\
\label{tmom}
\end{eqnarray}
where $\pi^i$ ($\pi_T$) is conjugate to $A_i$ ($T$), and $\pi^i$ satisfies
the Gauss constraint $\partial_i \pi^i = 0$.
The Hamiltonian is obtained by the following Legendre transformation
\begin{equation}
{\cal H} = \pi^i E_i + \pi_T \dot T - {\cal L} + \pi^i \partial_i A_0.
\end{equation}
{}From now on, let us use matrix notation for simplicity.
Then the quantities which we have defined are denoted by in matrix
forms in temporal gauge $A_0 = 0$,
\begin{eqnarray}\label{matrx}
X &=& X_{00} \det X_{ij} + E^+ D E^-, \label{matrx1} \\
\pi &=& \frac{V}{\sqrt{-X}}\frac{E^+ D + D E^-}{2}, \label{matrx2} \\
\pi_T &=& \frac{V}{\sqrt{-X}} \left( \dot T \det X_{ij}
- \frac{E^+ D \partial T - \partial T D E^-}{2} \right),\label{matrx3} \\
{\cal H} &=& \frac{V}{\sqrt{-X}}\left(\frac{E^+ D E + E D E^-}{2}
             + \dot T^2 \det X_{ij} - \frac{E^+ D\partial T \dot T
             - \dot T\partial T D E^-}{2} - X\right)
\nonumber \\
         &=& \frac{V}{\sqrt{-X}} \left((N^2 - L h L )\det X_{ij}
             + \frac{E^+ D h L - L h D E^-}{2} \right),
\label{matrx4}
\end{eqnarray}
where all matrix indices are $i,j,k = 1, \cdots, p$.
Let us define a matrix for convenience in calculations
\begin{equation}\label{Xbar}
\bar X = h + F + \partial T \partial T,
\end{equation}
which has properties
\begin{equation}
D \bar X = \bar X D = \det X_{ij} I,
\end{equation}
where $I$ is $p\times p$ unit matrix.
Then we obtain the following relations
\begin{eqnarray} \label{rel}
\pi \bar X \pi &=& \pi h \pi + (\pi \partial T)^2,
\nonumber  \\
F\pi + \partial T \pi_T &=& \frac{V}{\sqrt{-X}} \left(
                        \dot T \partial T \det X_{ij}
                        - \frac{ E^+ - E^-}{2} \det X_{ij}
                        + \frac{E^+ D h - h D E^-}{2} \right).
\nonumber \\
\end{eqnarray}
Using the relations in Eq.~(\ref{rel}), we find
\begin{equation}\label{rel2}
\sqrt{\pi h \pi +  \pi_T^2 +  (F\pi + \partial T \pi_T)
h^{-1}(F\pi + \partial T \pi_T) + (\pi \partial T)^2 + V^2 \det X_{ij}}
= \frac{V}{\sqrt{-X} }N \det X_{ij},
\end{equation}
where the matrix $h^{-1}$ has components $(h^{-1})_{ij} = h^{ij}$.
Now we can rewrite the Hamiltonian representation using canonical
variables as
\begin{eqnarray}\label{gham}
{\cal H} &=& \frac{V}{\sqrt{-X}} N^2 \det X_{ij}
             - \pi F L + \pi_T \partial L
\nonumber \\
         &=& N \sqrt{\pi h \pi +  \pi_T^2
            + (F\pi + \partial T \pi_T) h^{-1}(F\pi + \partial T \pi_T)
             + (\pi \partial T)^2 + \pi_T^2 + V^2 \det X_{ij}}
\nonumber \\
         && \,\,- \pi F L + \pi_T \partial T L,
\end{eqnarray}
where we used the relation
\begin{equation}
E^+ D L h - L h D E^-
             = 2 \left(L h L \det X_{ij} - \frac{\sqrt{-X}}{V}
               \pi F L + \frac{\sqrt{-X}}{V} \pi_T \partial T L\right).
\end{equation}

Then the Hamiltonian equations in $V=0$ limit are given by
\begin{eqnarray}
\dot T &=& \frac{\partial {\cal H}}{\partial \pi_T}
              = \frac{N}{\sqrt{Y}}\left(\pi_T(1 + \partial T h^{-1} \partial T)
                + \partial T h^{-1} F \pi\right)
                + \partial T L,\\
\dot \pi_T &=& - \frac{\partial{\cal H}}{\partial T}
               = \partial_i\left[\frac{N}{\sqrt{Y}}
                 \left( \pi^i(\pi \partial T)
                 + \pi_T h^{ij}(F_{jk} \pi^k + \partial_j T \pi_T)\right) \right]
                 + \partial_i(\pi_T L^i),
\\
\dot A_i &=& \frac{\partial {\cal H}}{\partial \pi^i}
                  = \frac{N}{\sqrt{Y}} \left( h \pi
                    + \partial T(\pi\partial T)
                    - F h^{-1} (F\pi + \partial T \pi_T)\right)_i
                    - (FL)_i,
\\
\dot \pi^i &=& -\frac{\partial {\cal H}}{\partial A_i}
                    = \partial_j\left(\frac{N}{\sqrt{Y}} (\pi^i h^{jk}
                      - \pi^j h^{ik})(F\pi + \partial T \pi_T)_k\right)
                      -\partial_j(\pi^j L^i - \pi^i L^j),
\nonumber \\
\end{eqnarray}
where $Y\equiv \pi h \pi + \pi_T^2  + (F\pi + \partial T \pi_T)
h^{-1}(F\pi + \partial T \pi_T) + (\pi \partial T)^2 $.

\section{Second Order Perturbation Equations}

General perturbation equations for Eqs.~(\ref{ab1}), (\ref{const1})
are given by,
\begin{eqnarray}
&& \partial_- \vec a^{(n)}
        = -\frac{\partial_y \vec n^{(0)}}{2 n_{0x}} b^{(n)}_y
          -\frac1{2 n_{0x}} \sum^{n-1}_{i=1}
           (\vec b^{(n-i)}\cdot\vec\partial)\vec a^{(i)},
\label{pert1} \\
&& \partial_+ \vec b^{(n)}
        = - \frac{\partial_y \vec n^{(0)}}{2 n_{0x}} a^{(n)}_y
          - \frac1{2 n_{0x}}\sum^{n-1}_{i=1}
            (\vec a^{(n-i)}\cdot\vec\partial)\vec b^{(i)},
\label{pert2} \\
&&2 \vec n^{(0)}\cdot\vec a^{(n)} = - \sum^{n-1}_{i=1} \vec a^{(i)}\cdot
                               \vec a^{(n-i)},
\label{pert3} \\
&&2 \vec n^{(0)}\cdot\vec b^{(n)} = - \sum^{n-1}_{i=1} \vec b^{(i)}\cdot
                               \vec b^{(n-i)}.
\label{pert4}
\end{eqnarray}

Now let us consider the second order perturbation equations.
The $y$-components of the second order perturbation equations
in Eqs.~(\ref{pert1}), (\ref{pert2}) are written by
\begin{eqnarray}
\partial_- a^{(2)}_y &=& -\frac1{2 n_{0x}}
                        (\vec b^{(1)}\cdot \vec\partial) a^{(1)}_y
\nonumber \\
                     &=& \left(-\frac1{2n_{0x}} \tilde g +
                          \frac{n_{0x}'}{(2 n_{0x})^2}
                          F\right) f^{(1,0)}
                         -\frac{n_{0x}'}{2 n_{0x}} t
                          \tilde ff^{(1,0)}
                         -\frac1{2n_{0x}} \tilde f f^{(0,1)},
\label{sod1} \\
\partial_+ b^{(2)}_y &=& -\frac1{2 n_{0x}}
                        (\vec a^{(1)}\cdot \vec\partial) b^{(1)}_y
\nonumber \\
                     &=& \left(-\frac1{2n_{0x}} g +
                          \frac{n_{0x}'}{(2 n_{0x})^2}
                          \tilde F\right) \tilde f^{(1,0)}
                         + \frac{n_{0x}'}{2 n_{0x}} t
                           f \tilde f^{(1,0)}
                         -\frac1{2n_{0x}}  f \tilde f^{(0,1)},
\label{sod2}
\end{eqnarray}
where we used the results of the first order perturbation
(\ref{fods1}) and (\ref{fods2}).
Then we obtain the solutions of these inhomogeneous first order
differential equations,
\begin{eqnarray}
a^{(2)}_y &=& h(x^+,y) -\left(\frac1{2n_{0x}} \tilde G
            + \frac{n_{0x}'}{2 n_{0x}} t
            \left(F + \tilde F \right)
             + \frac{n_{0x}'}{(2 n_{0x})^2}
             \tilde {\cal F}\right) f^{(1,0)}
          - \frac1{2n_{0x}} \tilde F f^{(0,1)},
\nonumber \\
\label{sods1} \\
b^{(2)}_y &=& \tilde h(x^-,y) -\left(\frac1{2n_{0x}}  G
            - \frac{n_{0x}'}{2 n_{0x}} t
            \left(F + \tilde F \right)
             + \frac{n_{0x}'}{(2 n_{0x})^2}
              {\cal F}\right) \tilde f^{(1,0)}
            - \frac1{2n_{0x}}  F \tilde f^{(0,1)},
\nonumber \\
\label{sods2}
\end{eqnarray}
where $h$ and $\tilde h$ are arbitrary functions and
we define,
\begin{eqnarray}
G(x^+, y) &\equiv& \int^{x^+} dw ~g(w,y), \qquad
\tilde G(x^-, y) \equiv \int^{x^-} dw ~\tilde g(w,y),
\nonumber \\
{\cal F} (x^+, y) &\equiv& \int^{x^+} dw ~F(w,y), \qquad
\tilde {\cal F} (x^-, y) \equiv \int^{x^-} dw ~\tilde F(w,y).
\end{eqnarray}
Using the Eqs.~(\ref{fods1}), (\ref{fods2}), (\ref{sods1}),
we can obtain the second order perturbation equations for
the $x$-components of the Eqs.~(\ref{pert1}),(\ref{pert2})
as follows:
\begin{eqnarray}
\partial_- a^{(2)}_x &=& -\frac1{2 n_{0x}}
                          \left( b^{(2)}_y n_{0x}'
                          + (\vec b^{(1)}\cdot \vec \partial) a^{(1)}_x
                          \right)
\nonumber \\
                    &=& -\frac{n_{0x}'}{2n_{0x}} \tilde h
                        + \frac{n_{0x}'}{(2 n_{0x})^2}
                         F \tilde f^{(0,1)}
                         + \left(-\frac1{2n_{0x}}\tilde g + \frac{\partial_y
                         n_{0x}}{(2 n_{0x})^2}F -\frac{n_{0x}'}{2 n_{0x}}
                          t \tilde f \right) g^{(1,0)}
\nonumber \\
                    & & + \left(\frac{n_{0x}'}{
                         (2 n_{0x})^2} G
                       + \frac{(n_{0x}')^2}{(2 n_{0x})^3}
                        {\cal F}
                       -\frac{(n_{0x}')^2}{(2 n_{0x})^2} t
                       \left(F + \tilde F\right)\right) \tilde f^{(1,0)}
\nonumber \\
                 & & + \left(\frac{n_{0x}'}{(2 n_{0x})^2} \tilde g
                    - \frac{(n_{0x}')^2}{(2 n_{0x})^3} F
                    - \frac1{2 n_{0x}} g^{(0,1)}
                    + \frac{n_{0x}'}{(2 n_{0x})^2}
                      \tilde F^{(0,1)}\right.
\nonumber \\
            & &~~  \left. + \frac{2\left(n_{0x} ( \partial_y^2 n_{0x})
                  -(n_{0x}' )^2\right)}{ (2n_{0x})^3} \tilde F
                - \frac{(n_{0x}')^2}{(2 n_{0x})^2}t\tilde f
                \right)\tilde f, \\
\partial_+ b^{(2)}_x &=& -\frac1{2 n_{0x}}
                          \left( a^{(2)}_y n_{0x}'
                          + (\vec a^{(1)}\cdot \vec \partial) b^{(1)}_x
                          \right)
\nonumber \\
                    &=& -\frac{n_{0x}'}{2n_{0x}} h
                        + \frac{n_{0x}'}{(2 n_{0x})^2}
                          \tilde F f^{(0,1)}
                        + \left(-\frac1{2n_{0x}} g + \frac{\partial_y
                          n_{0x}}{(2 n_{0x})^2}\tilde F
                          +\frac{n_{0x}'}{2 n_{0x}}
                          t  f \right) \tilde g^{(1,0)}
\nonumber \\
                    & & + \left(\frac{n_{0x}'}{
                         (2 n_{0x})^2} \tilde G
                       + \frac{(n_{0x}')^2}{(2 n_{0x})^3}
                        \tilde {\cal F}
                       + \frac{(n_{0x}')^2}{(2 n_{0x})^2} t
                       \left(F + \tilde F\right)\right)  f^{(1,0)}
\nonumber \\
    & & + \left(\frac{n_{0x}'}{(2 n_{0x})^2}  g
                - \frac{(n_{0x}')^2}{(2 n_{0x})^3} \tilde F
                - \frac1{2 n_{0x}} \tilde g^{(0,1)}
                + \frac{n_{0x}'}{(2 n_{0x})^2}
                   F^{(0,1)}\right.
\nonumber \\
            & &~~  \left. + \frac{2\left(n_{0x} ( \partial_y^2 n_{0x})
                  -(n_{0x}' )^2\right)}{(2n_{0x})^3}  F
                + \frac{(n_{0x}')^2}{(2 n_{0x})^2}t f
                \right)f.
\end{eqnarray}
\noindent
Using these results, we find
\begin{eqnarray}
\partial_+ \partial_- a^{(2)}_x
                &=& \left(\frac{n_{0x}'}{(2 n_{0x})^2}g
                   - \frac{(n_{0x}')^2}{(2 n_{0x})^3}
                           \tilde F
                   - \frac{(n_{0x}')^2}{(2 n_{0x})^2}t f
                       \right)\tilde f^{(1,0)}
\nonumber \\
             & & + \frac{n_{0x}'}{(2 n_{0x})^2}
                   f \tilde f^{(0,1)}
                 + \frac{n_{0x}'}{(2 n_{0x})^2}
                   (f - \tilde f) g^{(1,0)}
                 -  \frac1{2 n_{0x}}\tilde f g^{(1,1)}
\nonumber \\
            & & + \left( - \frac1{2 n_{0x}} \tilde g
                    +\frac{n_{0x}'}{(2 n_{0x})^2}F
              - \frac{n_{0x}'}{2 n_{0x}}t\tilde f \right) g^{(2,0)}
              - \frac{(n_{0x}')^2}{(2 n_{0x})^3}
                   ( f + \tilde f)\tilde f,
\nonumber \\  \label{sod3} \\
\partial_+ \partial_- b^{(2)}_x
                &=& \left(\frac{n_{0x}'}{(2 n_{0x})^2}
                    \tilde g
                   - \frac{(n_{0x}')^2}{(2 n_{0x})^3} F
                   + \frac{(n_{0x}')^2}{(2 n_{0x})^2}
                     t \tilde f \right) f^{(1,0)}
\nonumber \\
             & & + \frac{n_{0x}'}{(2 n_{0x})^2}
                   \tilde f f^{(0,1)}
                 - \frac{n_{0x}'}{(2 n_{0x})^2}
                   (f - \tilde f) \tilde g^{(1,0)}
                  -  \frac1{2 n_{0x}} f \tilde g^{(1,1)}
\nonumber \\
            & & + \left( - \frac1{2 n_{0x}}  g
                    +\frac{n_{0x}'}{(2 n_{0x})^2}\tilde F
                    + \frac{n_{0x}'}{2 n_{0x}}t  f
                 \right)\tilde g^{(2,0)}
              - \frac{(n_{0x}')^2}{(2 n_{0x})^3}
                   ( f + \tilde f) f,
\nonumber \\  \label{sod4} \\
\partial_+\partial_- a_y^{(2)}
           &=& \frac{n_{0x}'}{(2 n_{0x})^2}
           (f-\tilde f) f^{(1,0)}
           -\frac{1}{2 n_{0x}} \tilde f f^{(1,1)}
\nonumber \\
          & & -\left(\frac{1}{2 n_{0x}}\tilde g
              + \frac{n_{0x}'}{2 n_{0x}}t\tilde f
              -\frac{n_{0x}'}{(2 n_{0x})^2} F
              \right ) f^{(2,0)},
 \label{sod5} \\
\partial_+\partial_- b_y^{(2)}
          &=&\frac{n_{0x}'}{(2 n_{0x})^2}
           (\tilde f- f) \tilde f^{(1,0)}
           -\frac{1}{2 n_{0x}}  f \tilde f^{(1,1)}
\nonumber \\
           & & -\left(\frac{1}{2 n_{0x}} g
              - \frac{n_{0x}'}{2 n_{0x}}t f
              -\frac{n_{0x}'}{(2 n_{0x})^2} \tilde F
              \right ) \tilde f^{(2,0)}.
\label{sod6}
\end{eqnarray}

\end{document}